\documentclass[a4paper,11pt]{article}%
\pdfoutput=1 
\usepackage{jheppub}
\usepackage{amsmath}
\usepackage{amsfonts}
\usepackage{amssymb}
\usepackage{graphicx}%
\usepackage[T1]{fontenc} 

\newcommand{\dhd}{{\textstyle d}
\lower.03ex\hbox{\kern-0.40em$^{\scriptstyle-}$}\kern-0.08em{}}  

\newcommand{\dbar}{{\textstyle \delta}
\lower.03ex\hbox{\kern-0.38em$^{\scriptstyle-}$}\kern-0.05em{}}
\newcommand{\half}{{1\over 2}}
\newcommand{\bu}{{\bullet}}

\newcommand{\bsi}{{\bar \psi}}

\newcommand{\cald}{{\cal D}}  
\newcommand{\calf}{{\cal F}} 
 
\newcommand{\calh}{{\cal H}} 
\newcommand{\cali}{{\cal I}}     
  
\newcommand{\calo}{{\cal O}}

\newcommand{\hatH}{{\hat H}}

 \newcommand{\vecw}{{\vec w}} 
\newcommand{\vecx}{{\vec x}}
 \newcommand{\vecy}{{\vec y}}

\newcommand{\tilj}{{\tilde j}}

\newcommand{\tipsi}{{\tilde\psi}}

\newcommand{\tilcaf}{\tilde{\cal F}} 
\newcommand{\ticalo}{\tilde{\cal O}} 
\newcommand{\ticalf}{\tilde{\cal F}}

\newcommand{\tilA}{\tilde{A}} 
\newcommand{\tilD}{\tilde{D}} 
 
\newcommand{\tilF}{\tilde{F}} 
 
\newcommand{\tilT}{\tilde{T}} 
\newcommand{\tilU}{\tilde{U}}

\newcommand{\calr}{{\cal R}}

\pdfoutput=1
\abstract{We study the rapidity evolution of gluon transverse momentum dependent distributions
appearing in processes of particle production and show how this evolution
changes from small to moderate Bjorken $x$.}
\arxivnumber{}
\affiliation{$^{*}$ Physics Dept., Old Dominion University, Norfolk VA 23529,USA
and Theory Group, JLAB, 12000 Jefferson Ave, Newport News, VA 23606,USA}
\affiliation{$^{\dag}$ Theory Group, JLAB, 12000 Jefferson Ave, Newport News, VA 23606,USA}
\emailAdd{balitsky@jlab.org}
\emailAdd{atarasov@jlab.org}
\begin{document}

\title{\boldmath Gluon TMD in particle production from low to moderate $x$}
\author{I. Balitsky$^{*}$ and A. Tarasov$^{\dag}$}
\preprint{JLAB-THY-16-2229}
\maketitle

\flushbottom

\section{Introduction\label{aba:sec1}}

\bigskip

The TMDs \cite{cs1, jimayuan,collinsbook} (also called unintegrated parton distributions) are widely used in the analysis of various scattering processes like SIDIS or Drell-Yan. The TMD generalizes the usual concept of parton density by allowing PDFs to
depend on intrinsic transverse momenta in addition to the usual longitudinal momentum fraction variable. 
At low energies the relevant quantities are quark TMDs and there is a vast literature on the application of quark TMDs 
for analysis of cross sections of processes measured at JLab and elsewhere (see e.g. Refs. \cite{ex1, ex2, ex3, ex4, ex5, ex6, ex7, ex8, ex9, ex10, ex11, ex12}, for review see also Refs. \cite{ex13, ex14, ex15}).
However, since at the future EIC accelerator the majority of the produced particles will be gluons  one needs to study also the evolution of gluon TMDs. Moreover, the EIC energies may be in the intermediate region between hard physics described by linear CSS evolution \cite{cs2}  and low-x physics described by non-linear BK/JIMWLK evolution \cite{npb96,yura,jimwalk} so one needs to study the transition of the evolution of gluon TMDs  between these two regimes.\footnote{For the study of quark TMDs in the small-x regime, see Ref. \cite{kovsievert}.}

The gluon TMD (unintegrated gluon distribution) is defined as \cite{muldrod}
\begin{eqnarray}
&&\hspace{-0mm}
\cald(x_B,k_\perp,\eta)~=~\!\int\!d^2z_\perp~e^{i(k,z)_\perp}\cald(x_B,z_\perp,\eta),
\label{gTMD}\\
&&\hspace{-0mm}
\alpha_s\cald(x_B,z_\perp,\eta)~=~{-x_B^{-1}\over 8\pi^2(p\cdot n)}\!\int\! du ~e^{-ix_Bu(pn)} \langle P|\calf^a_\xi(z_\perp+un)
[z_\perp-\infty n,-\infty n]^{ab}\calf^{b\xi}(0)|P\rangle
\nonumber
\end{eqnarray}
where $|P\rangle$ is an unpolarized target with momentum $p$ (typically proton) and $n$ is a light-like vector.
Hereafter we use the notation
\begin{eqnarray}
&&\hspace{-0mm}
\calf^a_\xi(z_\perp+un)~\equiv~n^\mu gF^m_{\mu \xi}(un+z_\perp)[u n+z_\perp,-\infty n+z_\perp]^{ma}
\label{1.2}
\end{eqnarray}
where $[x,y]$ denotes straight-line gauge link connecting points $x$ and $y$:
\begin{equation}
~[x,y]~\equiv~{\rm P}e^{ig\int\! du~(x-y)^\mu A_\mu(ux+(1-u)y)}
\label{defu}
\end{equation}
There are more involved definitions with Eq. (\ref{gTMD}) multiplied by some Wilson-line factors \cite{collinsbook, echevidsci} 
following from CSS factorization \cite{cs2} but we will discuss the ``primordial'' TMD (\ref{gTMD}).

It is well known, however, that gluon TMDs are not universal in a sense that the direction of gauge links providing gauge invariance
depends on the type of processes under consideration, see Ref. \cite{collins1}.
For example, TMDs entering the description of processes particle production have light-like gauge links starting at minus infinity as in  Eq. (\ref{1.2}), but  
TMDs which appear in the analysis of semi-inclusive processes have gauge links stretching to plus infinity 
(so the corresponding expression for TMDs is obtained  by replacement $-\infty\leftrightarrow\infty$ in  Eq.  (\ref{1.2})). For a more complicated processes the structure of gauge links may be even more involved, see e.g. Ref. \cite{bmmuld}.

In our recent paper \cite{gTMD1} we have obtained the leading-order evolution equation for gluon TMDs for semi-inclusive processes like semi-inclusive deep inelastic scattering (SIDIS). The obtained equation describes the rapidity evolution of gluon TMDs in the whole region of small to moderate Bjorken $x_B$ and for any transverse momentum. It interpolates between the linear DGLAP and Sudakov evolution equations at moderate $x_B$ and the non-linear BK equation for small $x_B$. In this paper we extend our analysis to the case of gluon TMDs
appearing in particle production processes with gauge links extending to minus infinity in the light-cone (LC) time direction. The analysis is very close to the study of our paper \cite{gTMD1} so we will streamline the presentation of technical details paying attention to  differences between these two cases (with links going to plus or minus infinity).  The final evolution equations
are similar (but in general not the same!)  to TMDs with gauge links extending to plus infinity.

 The paper is organized as follows. In Sec. 2 we remind the  logic of rapidity factorization 
 for the inclusive particle production and rapidity evolution.  
 In Sec. 3 we discuss rapidity evolution of gluon TMDs and calculate the leading-order kernel of the evolution equation. 
 We present the final form of the evolution equation in Sect. 4 and discuss BK, Sudakov and DGLAP limits in Sect. 5  
 and linearized equation in Sect. 6. Sect. 7 contains conclusions and outlook. 
 The necessary formulas for propagators near the light cone and in the shock-wave
 background can be found in  Appendices. 

\section{TMDs in particle production}

To simplify the description of particle production, let us consider the model where a (colorless) scalar particle 
can be produced by gluon-gluon fusion through the vertex coming from the Lagrangian
\begin{equation}
S_\Phi~=~\lambda\! \int\! d^4z ~F_{\mu\nu}^a(z)F^{a\mu\nu}(z)\Phi(z)
\end{equation}
One may consider this as a model of Higgs production by gluon fusion in the region where transverse momentum 
of produced Higgs boson is smaller than the mass of the top quark.

Let us consider the production of this $\Phi$-boson in the high-energy scattering of a virtual photon with virtuality $\sim$ few GeV off the hadron target.
As demonstrated in the Appendix \ref{appcrsc}  the cross section of $\Phi$-boson production can be represented by a double
functional integral 
\begin{eqnarray}
&&\hspace{-2mm}
\sigma_{\mu\nu}~
=~{\lambda^2\over 2\pi}\!\int \! d^4w d^4x d^4y e^{iqw-ikx+iky}
\int^{\tilA(t_f)=A(t_f)}\! D\tilA D\tilde{\bar\psi}D\tilde{\psi}DA D\bsi D\psi ~
\label{dabl}\\
&&\hspace{-1mm}
\times~\Psi^\ast_p(\vec{\tilA}(t_i),\tipsi(t_i))e^{-iS_{\rm QCD}(\tilA,\tipsi)}e^{iS_{\rm QCD}(A,\psi)}
\tilj_\mu(w)\tilF^2(x)F^2(y)j_\nu(0)\Psi_p(\vec{A}(t_i),\psi(t_i))
\nonumber
\end{eqnarray}
where $\Psi_p$ are proton wave functionals at the initial time $t_i\rightarrow -\infty$.
(The boundary condition $\tilA(\vec{x},t_f\rightarrow\infty)=A(\vec{x},t_f\rightarrow\infty)$ and similar condition for quark fields  reflects the sum over all intermediate states $X$).

We will analyze the energy dependence of this cross section using the high-energy OPE in Wilson lines. 
To this end, we integrate over rapidities greater than the rapidity of the produced $\Phi$-boson  
$Y>\eta_\phi$ and leave the fields with $Y<\eta_\phi$  to be integrated over later. 
The result of the integration over $Y>\eta_\phi$ is the coefficient function (called ``impact factor'') in
front of the Wilson-line operator(s) made of gluons (and quarks) with rapidities $Y<\eta_\phi$.
 (Strictly speaking, we integrate over rapidities $Y>\eta_\phi-\epsilon$ so the vertex of $\Phi$-boson production is included into the impact factor).
To make connections with parton model we will have in mind the frame where target's velocity is large and call the small $\alpha$ fields by the name ``fast fields'' and large $\alpha$ fields by 
``slow'' fields.  Of course, ``fast'' {\it vs} ``slow'' depends on frame but we will stick to naming fields as they appear 
in the projectile's frame. (Note that in Ref. \cite{npb96} the terminology is opposite, as appears in the target's frame).
As discussed in Ref. \cite{npb96},  the interaction of ``slow'' gluons of large $Y$ with ``fast'' fields of 
small $Y$  is described by eikonal  gauge factors and the integration over slow fields  results in  
Feynman diagrams in the background of fast fields which form a thin shock wave due to Lorentz contraction.
\footnote{ An exceptional case discussed later is when the transverse momenta of the external field are much smaller than
than the characteristic transverse momenta in the impact factor. In this case the ``shock wave'' is no longer narrow and one 
needs the light-cone approximation rather than the shock-wave one. However, if the virtuality of the photon is $\sim$ few GeV 
the characterisic transverse momenta of the  impact factor and of the fast ``external fields''  are of the same order of magnitude so the shock-wave approximation is applicable.} 
In the spirit of high-energy OPE, the rapidity of the gluons is restricted from above by the ``rapidity divide'' $\eta$ separating the impact factor and the matrix element so the proper definition of $U_x$ is 
\begin{eqnarray}
&&\hspace{-0mm} 
 U^\eta_x~=~{\rm Pexp}\Big[ig\!\int_{-\infty}^\infty\!\! du ~p_1^\mu A^\eta_\mu(up_1+x_\perp)\Big],
\nonumber\\
&&\hspace{-0mm} 
A^\eta_\mu(x)~=~\int\!{d^4 k\over 16\pi^4} ~\theta(e^\eta-|\alpha|)e^{-ik\cdot x} A_\mu(k)
\label{cutoff}
\end{eqnarray}
where  the  Sudakov variable $\alpha$ is defined as usual,  $k=\alpha p_1+\beta p_2+k_\perp$.
We define the light-like vectors $p_1$ and $p_2$ close to projectile and target's momenta $q$ and $p$ so that $q=p_1+{q^2\over s}p_2$  and $p=p_2+{m^2\over s}p_1$. We use metric $g^{\mu\nu}~=~(1,-1,-1,-1)$ so 
that $p\cdot q~=~(\alpha_p\beta_q+\alpha_q\beta_p){s\over 2}-(p,q)_\perp$. For the coordinates we use 
the notations $x_\bu\equiv x_\mu p_1^\mu$ and $x_\ast\equiv x_\mu p_2^\mu$ for dimensionless light-cone coordinates ($x_\ast=\sqrt{s\over 2}x_+$ and $x_\bu=\sqrt{s\over 2}x_-$).

In accordance with general background-field formalism we separate the gluon field into the ``classical'' background part and
``quantum'' part
$$
A_\mu~\rightarrow~A^{\rm cl}_\mu+A^{\rm q}_\mu,~~~~~\psi~\rightarrow~\psi^{\rm cl}+\psi^{\rm q}
$$
where the ``classical'' fields are fast $(\alpha<\sigma=e^\eta)$ and ``quantum'' fields are slow  $(\alpha>\sigma=e^\eta)$.
It should be emphasized that our ``classical'' field does not satisfy the equation $D^\mu F^{cl}_{\mu\nu}=0$; rather, 
$(D^\mu F^{\rm cl}_{\mu\nu})^a=-g\bsi\gamma_\nu t^a\psi$ where $\psi$ are the ``classical'' (i.e. fast) quark fields. 

The first-order term in the expansion of the operator $F^m_{\bu i}(y_\ast,y_\perp)[y_\ast, -\infty]_y^{ma}$ in quantum fields has the form
\begin{eqnarray}
&&\hspace{-11mm}
F_{\bu i}^m(y_\ast,y_\perp)[y_\ast,-\infty]_y^{ma}~\stackrel{\rm 1st}{=}~
{s\over 2}{\partial\over\partial y_\ast}A_i^{m{\rm q}}(y_\ast,y_\perp)[y_\ast,-\infty]_y^{ma}
\label{3.3}\\
&&\hspace{-11mm}
-~\partial_iA_\bu^{m{\rm q}}(y_\ast,y_\perp)[y_\ast,-\infty]_y^{ma}
+i\!\int_{-\infty}^{y_\ast}\! d{2\over s}z'_\ast ~F_{\bu i}^m(y_\ast,y_\perp)([y_\ast,z'_\ast]_yA_\bu^{\rm q}(z'_\ast,y_\perp)[z'_\ast,-\infty]_y)^{ma}
\nonumber
\end{eqnarray}
(to save space, we omit the label ${}^{\rm cl}$ from classical fields). 

\begin{figure}[htb]
\begin{center}
\includegraphics[width=50mm]{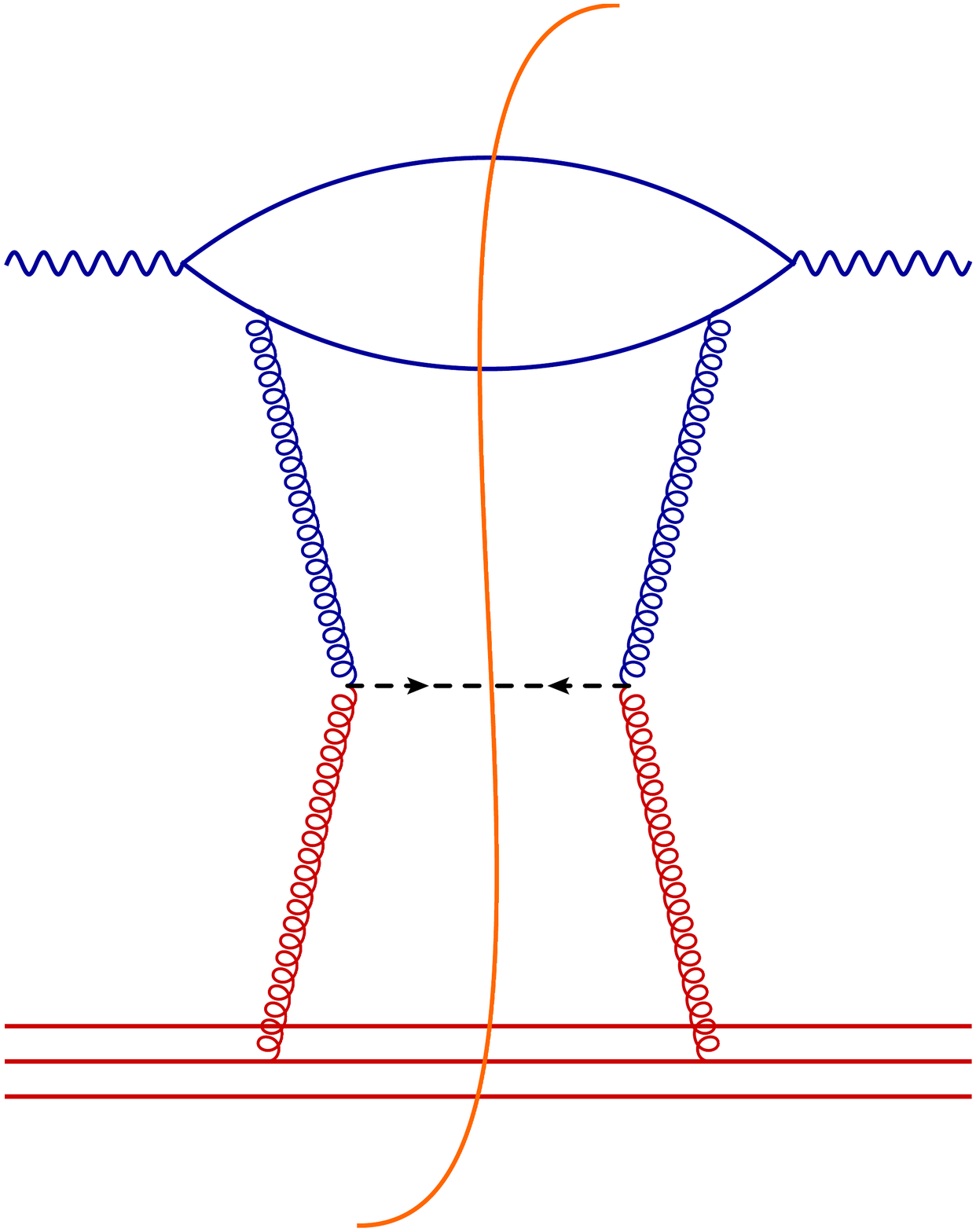}
\hspace{2mm}
\includegraphics[width=10mm]{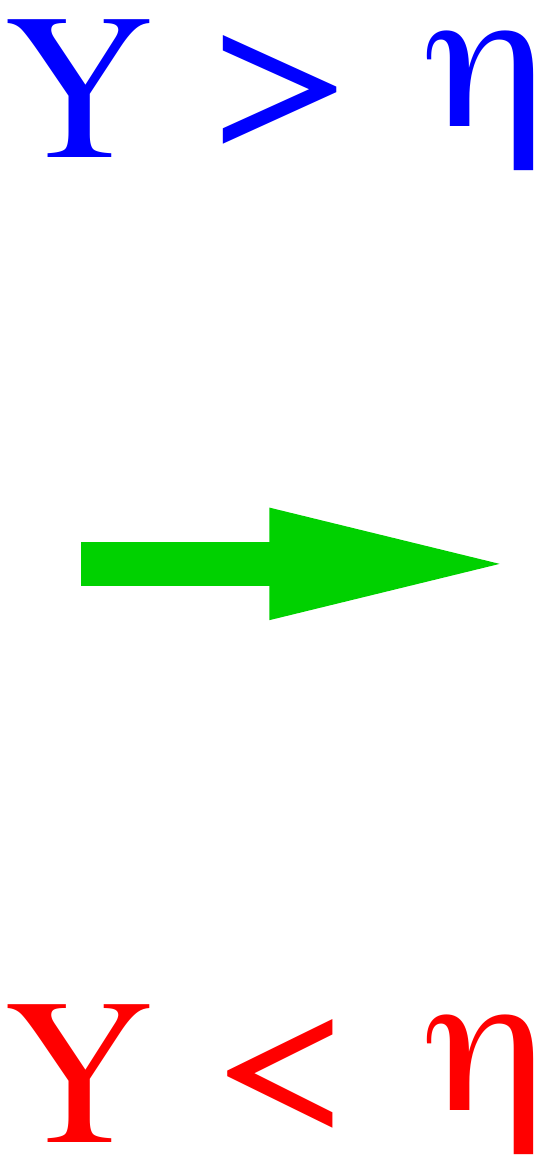}
\hspace{2mm}
\includegraphics[width=60mm]{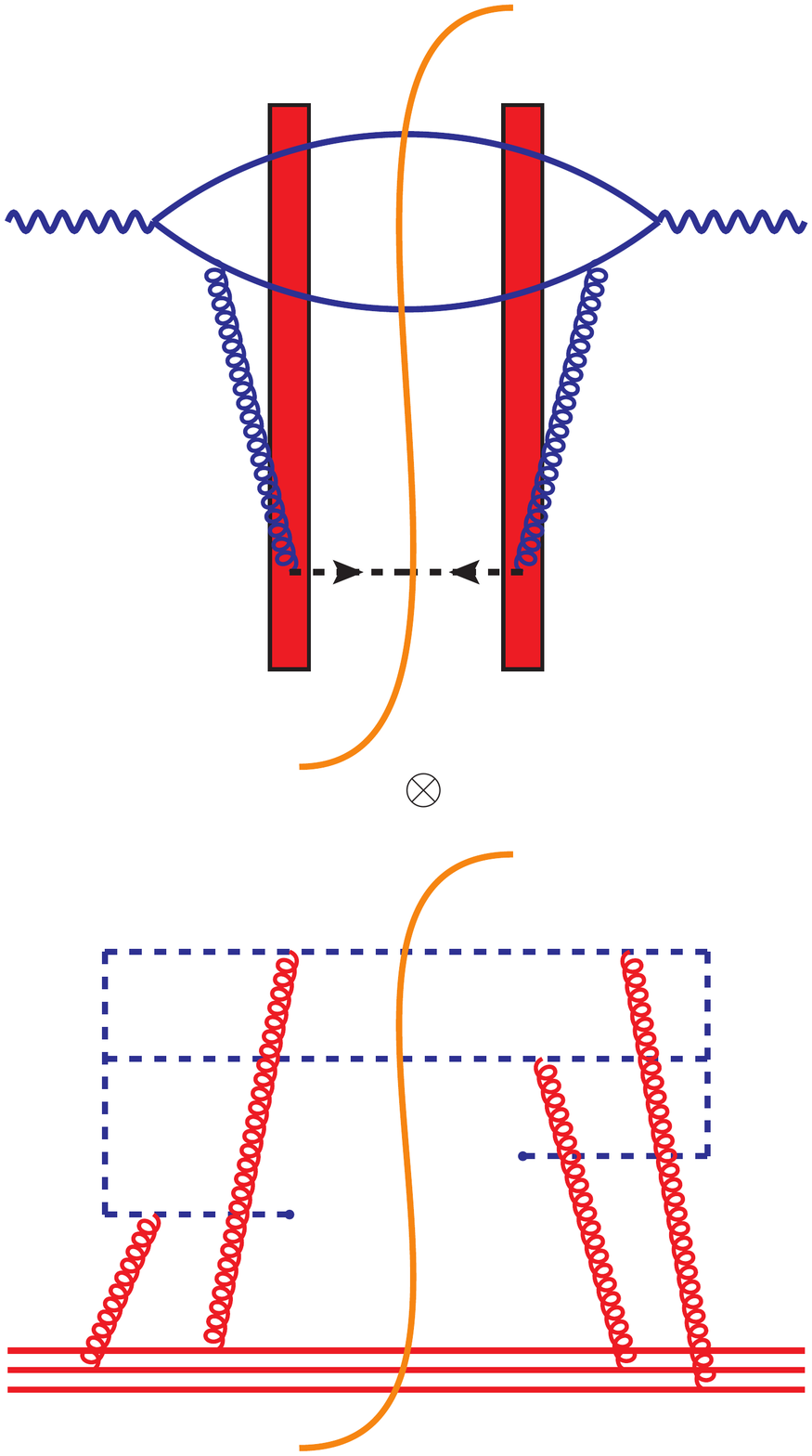}
\end{center}
\caption{Rapidity factorization for particle production. The dashed lines denote gauge links.\label{fig:1}}
\end{figure}

In the leading order the impact factor is given by the diagram shown in Fig. \ref{fig:1}. 
The quark propagator in the external field has the form
\begin{eqnarray}
&&\hspace{-11mm}
\langle \tilde{\psi}(x)\bar{\psi}(y)\rangle
\label{kvprop}\\
&&\hspace{-11mm}
\stackrel{x_\ast,y_\ast<0}{=}~\int_\sigma^\infty\!{\dhd\alpha\over 2\alpha^2s}~e^{-i\alpha(x-y)_\bu}
(x_\perp|e^{-i{p_\perp^2\over\alpha s}x_\ast}(\alpha\!\not\! p_1+\not\! p_\perp)
\not\! p_2\tilU^\dagger U(\alpha\!\not\! p_1+\not\! p_\perp)
e^{i{p_\perp^2\over\alpha s}y_\ast}|y_\perp)
\nonumber
\end{eqnarray}
where $\sigma=e^\eta$ is the lower rapidity cutoff for the impact factor (and upper cutoff for $\alpha$'s in Wilson lines). Hereafter we use Schwinger's notations
\begin{equation}
(x_\perp|f(p_\perp)|y_\perp)~\equiv~ \int\! \dhd^2p_\perp~e^{i(p,x-y)_\perp}f(p_\perp), ~~~~~(x_\perp|p_\perp)~=~e^{i(p,x)_\perp}
\label{schwingerperp}
\end{equation}
Note that unlike the case of total cross section, here we consider particle production so the gluon lines in Fig. \ref{fig:1} terminate 
at the $\Phi$-boson emission point leading to gluon TMDs rather than proper Wilson lines (stretching from minus to plus infinity in LC-time direction). 
Indeed, the gluon propagator with one point in the shock wave has the form of the free propagator multiplied by the gauge link going
from point $y$ to $-\infty$ in the $p_1$ direction \cite{gTMD1}:
\begin{eqnarray}
&&\hspace{-11mm}
\langle A_\mu(z)F_{\ast j}(y)\rangle~=~\frac{i}{2}\!\int_\sigma^\infty\!{\dhd\alpha\over 2\alpha}e^{-i\alpha(y-z)_\bu}
(z_\perp|e^{-i{p_\perp^2\over\alpha s}(y-z)_\ast}(\alpha s g^\perp_{\mu j}+2p_{2\mu}p_j)|y_\perp)[-\infty, y_\ast]_y
\label{impprop1}
\end{eqnarray}
Since the propagators (\ref{kvprop}) and (\ref{impprop1}) have simple structure one can calculate the integrals in Fig. \ref{fig:1} 
and  the result has the form
\begin{eqnarray}
&&\hspace{-1mm}
\lambda^{-2}\sigma_{\mu\nu}~=~
\label{dabresult1}\\
&&\hspace{-1mm}
=~\int\! d^2z_{1_\perp} d^2z_{2_\perp}d^2x_\perp d^2y_\perp ~
I_{\mu\nu}^{ij}(z_{1_\perp},z_{2_\perp},x_\perp,y_\perp;\eta)e^{i(k,x-y)_\perp}\!\int\! dx_\ast dy_\ast~e^{-i\beta_B(x_\ast-y_\ast)}
\nonumber\\
&&\hspace{-1mm}
\times\int^{\tilA(\infty)=A(\infty)}\! D\tilA D\tilde{\bar\psi}D\tilde{\psi}DA D\bsi D\psi ~
\Psi^*_p(\tilA,\tipsi)|_{t_i=-\infty}~e^{-iS_{\rm QCD}(\tilA,\tipsi)}
e^{iS_{\rm QCD}(A,\psi)}
\nonumber\\
&&\hspace{-1mm}
\times{\rm tr}\{\tilU_{z_2}[z_{2_\perp},x_\perp]_{-\infty}[-\infty_\ast,x_\ast]_x
\tilF_{\bu i}(x_\ast,x_\perp)[x_\ast,-\infty_\ast]_x[x_\perp,z_{1_\perp}]_{-\infty}\tilU_{z_1}^\dagger
\nonumber\\
&&\hspace{-1mm}
\times~U_{z_1}[z_{1_\perp},y_\perp]_{-\infty}[-\infty_\ast,y_\ast]_y
F_{\bu j}(y_\ast,y_\perp)[y_\ast,-\infty_\ast]_y[y_\perp,z_{2_\perp}]_{-\infty}U_{z_2}^\dagger\}
\Psi_p(A,\psi)|_{t_i=-\infty}
\nonumber
\end{eqnarray}
where ${\rm tr}\{...\}$ is the color trace in the fundamental representation and $I^{ij}_{\mu\nu}(z_{1_\perp},z_{2_\perp},x_\perp,y_\perp;\sigma)$ is the impact factor with the 
lower rapidity cutoff $\eta=\ln\sigma$.\footnote{Both impact factor and matrix element of Wilson-line operators depend on the ``rapidity divide'' $\sigma$ but this dependence is canceled in their product.}
 Hereafter we use the short-hand notations for gauge links
\begin{equation}
[x_\ast,z_\ast]_x~\equiv~[{2\over s}x_\ast p_1+x_\perp,{2\over s}z_\ast p_1+x_\perp]
\end{equation}
and
\begin{equation}
[x_\perp,z_\perp]_{-\infty}~\equiv~[-{2\over s}\infty_\ast p_1+x_\perp,-{2\over s}\infty_\ast p_1+z_\perp]
\label{inftylink}
\end{equation}
As discussed in Ref. \cite{npb96}, the fast fields at lightcone time $\pm\infty$ are pure gauge so the precise
form of the contour in Eq. (\ref{inftylink}) is irrelevant.

The calculation of the impact factor $I^{ij}_{\mu\nu}(z_{1_\perp},z_{2_\perp},x_\perp,y_\perp;\eta)$ 
is similar  to the calculation of the NLO photon impact factor for the 
DIS structure functions carried out in Ref. \cite{nloif}. Since the explicit form of   $I^{ij}_{\mu\nu}$  is irrelevant for our purpose of 
finding the evolution of gluon TMDs and since in the real life the contribution  
of the diagram shown in Fig. \ref{fig:1} is a tiny correction to the total cross section of Higgs production in DIS we did not attempt to calculate this impact factor.  In the case of proton-proton scattering the impact factor 
should be given by another gluon TMD made of Wilson lines stretched in $p_2$ direction. We intend
to discuss the obtained factorization in a separate publication.

As demonstrated in Appendix \ref{appcrsc} (see Eq. (\ref{funtegral})), the double functional integral (\ref{dabresult1}) represents the matrix element
\begin{eqnarray}
&&\hspace{-1mm}
\lambda^{-2}\sigma_{\mu\nu}~=~
\nonumber\\
&&\hspace{-1mm}
=\int\! d^2z_{1_\perp} d^2z_{2_\perp}d^2x_\perp d^2y_\perp ~
I_{\mu\nu}^{ij}(z_{1_\perp},z_{2_\perp},x_\perp,y_\perp;\eta)e^{i(k,x-y)_\perp}\!\int\! dx_\ast dy_\ast~e^{-i\beta_B(x_\ast-y_\ast)}
\nonumber\\
&&\hspace{-1mm}
{\rm tr}\langle p|\tilT\{U_{z_2}[z_{2_\perp},x_\perp]_{-\infty}[-\infty_\ast,x_\ast]_x
F_{\bu i}(x_\ast,x_\perp)[x_\ast,-\infty_\ast]_x[x_\perp,z_{1_\perp}]_{-\infty}U_{z_1}^\dagger\}
\nonumber\\
&&\hspace{-1mm}
\times~T\{U_{z_1}[z_{1_\perp},y_\perp]_{-\infty}[-\infty_\ast,y_\ast]_y
F_{\bu j}(y_\ast,y_\perp)[y_\ast,-\infty_\ast]_y[y_\perp,z_{2_\perp}]_{-\infty}U_{z_2}^\dagger\}|p\rangle
\label{2.11}
\end{eqnarray}
Note that all the gluon operators in the r.h.s. of this equation are separated either by space-like or by 
light-like distances. In both cases, the operators commute 
\footnote{For the space-like separations this is trivial whereas
the commutation of operators on the light ray is proven in Ref. \cite{jaffe}.}
so one can erase $\tilT$ and $T$ signs and get the matrix element
\begin{eqnarray}
&&\hspace{-1mm}
{\rm tr}\langle p|[z_{2_\perp},x_\perp]_{-\infty}[-\infty_\ast,x_\ast]_x
F_{\bu i}(x_\ast,x_\perp)[x_\ast,-\infty_\ast]_x[x_\perp,z_{1_\perp}]_{-\infty}
\nonumber\\
&&\hspace{-1mm}
\times~[z_{1_\perp},y_\perp]_{-\infty}[-\infty_\ast,y_\ast]_y
F_{\bu j}(y_\ast,y_\perp)[y_\ast,-\infty_\ast]_y[y_\perp,z_{2_\perp}]_{-\infty}|p\rangle
\end{eqnarray}
Moreover, as we mentioned above, for the fast gluons the precise form 
of gauge link at infinity does not matter so we can connect points $x_\perp$ and $y_\perp$ by
a straight-line gauge link $[x_\perp,y_\perp]_{-\infty}$ (instead of 
$[x_\perp,z_{1_\perp}]_{-\infty}[z_{1_\perp},y_\perp]_{-\infty}$) and obtain the matrix
element 
\begin{eqnarray}
&&\hspace{-1mm}
{\rm tr}\langle p|[y_\perp,x_\perp]_{-\infty}[-\infty_\ast,x_\ast]_x
F_{\bu i}(x_\ast,x_\perp)[x_\ast,-\infty_\ast]_x
\nonumber\\
&&\hspace{-0mm}
\times[x_\perp,y_\perp]_{-\infty}
[-\infty_\ast,y_\ast]_y
F_{\bu j}(y_\ast,y_\perp)[y_\ast,-\infty_\ast]_y|p\rangle
\label{2.13}
\end{eqnarray}
proportional to gluon TMD (\ref{gTMD}). Note, however, that forward matrix element of this operator has an unbounded integration
over $x_\ast-y_\ast$. It is convenient to introduce the notation $\langle\!\langle  p|O|p\rangle\!\rangle$ for the forward matrix element of the operator $O$ stripped of 
this integration
\begin{eqnarray}
&&\hspace{-0mm} 
\langle p|\tilcaf^{a\eta}_i(\beta_B, z_\perp)\calf^{ai\eta}(\beta_B, 0_\perp)|p+\xi p_2\rangle 
\nonumber\\
&&\hspace{-0mm} 
=~2\pi\delta(\xi)\langle\!\langle  p|\tilcaf^{a\eta}_i(\beta_B, z_\perp)\calf^{ai\eta}(\beta_B, 0_\perp)|p\rangle\!\rangle
\label{2.14}
\end{eqnarray}
With this notation the unintegrated gluon TMD (\ref{gTMD}) can be represented as 
\begin{eqnarray}
&&\hspace{-0mm} 
\langle\!\langle p|\tilcaf^{a\eta}_i(\beta_B, z_\perp)\calf^{ai\eta}(\beta_B, 0_\perp)|p\rangle\!\rangle 
~=~-2\pi\beta_Bg^2\cald(\beta_B,z_\perp,\eta)
\label{TMD}
\end{eqnarray}
Returning to Eq. (\ref{2.13}),  since the dependence on $z_{i_\perp}$ is gone from the matrix element, we can integrate the impact factor over $z_{1_\perp}$ and $z_{2_\perp}$ and get the cross section as a convolution of the new impact factor $\cali_{\mu\nu}(x_\perp,y_\perp;\eta)$ with the gluon TMD
\begin{eqnarray}
&&\hspace{-1mm}
\lambda^{-2}\sigma_{\mu\nu}~
\label{TMDfakt}\\
&&\hspace{-1mm}=~
\int\! d^2x_\perp d^2y_\perp ~
\cali_{\mu\nu}^{ij}(x_\perp,y_\perp;\eta)e^{i(k,x-y)_\perp}
\langle\!\langle p|\ticalf^a_i(\beta_B, x_\perp)[x_\perp,y_\perp]_{-\infty}^{ab}\calf^b_j(\beta_B, y_\perp)|p\rangle\!\rangle ^\eta
\nonumber
\end{eqnarray}
where \footnote{Hereafter the notation $\tilcaf$ is just a reminder of different signs in the exponents of Fourier transforms 
in the definitions (\ref{calfs}).}
\begin{eqnarray}
&&\hspace{1mm}
\ticalf_i(\beta_B, x_\perp)~\equiv~{2\over s}\!\int\! dx_\ast~e^{-i\beta_Bx_\ast}\calf^a_i(x_\ast, x_\perp)
\nonumber\\
&&\hspace{1mm}
\calf_i(\beta_B, y_\perp)~\equiv~{2\over s}\!\int\! dy_\ast~e^{i\beta_By_\ast}\calf^a_i(y_\ast, y_\perp)
\label{calfs}
\end{eqnarray}
Note that the Wilson-line operators $U_z^\dagger$ and $U_z$ in Eq. (\ref{2.11}) cancel only 
when we take a sum over all
intermediate states. If we are interested in, say, production of another particle (at lower rapidity), we need 
to consider the full double functional integral (\ref{dabresult1}).

\section{Rapidity factorization and  evolution of TMDs in the leading order\label{sec2}}

We will study the rapidity evolution of the operator 
\begin{equation}
\tilcaf^{a\eta}_i(\beta_B, x_\perp)[x_\perp,y_\perp]_{-\infty}^{ab}\calf^{b\eta}_j(\beta_B, y_\perp)
\label{operator}
\end{equation}
Matrix elements of this operator between unpolarized hadrons can be parametrized as \cite{muldrod}
\begin{eqnarray}
&&\hspace{0mm} 
\int\! d^2z_\perp~e^{i(k,z)_\perp}\langle\!\langle p|\tilcaf^{a\eta}_i(\beta_B, z_\perp)\calf_j^{a\eta}(\beta_B, 0_\perp)|p\rangle\!\rangle^\eta 
~=~ \pi g^2\calr_{ij}(\beta_B, k_\perp;\eta)
\nonumber\\
&&\hspace{0mm}
\calr_{ij}(\beta_B, k_\perp;\eta)~=~-g_{ij}\beta_B\cald(\beta_B,k_\perp,\eta)
+\Big({2k_ik_j\over m^2}+g_{ij}{k_\perp^2\over m^2}\Big)\beta_B\calh(\beta_B,k_\perp,\eta)
\label{mael}
\end{eqnarray}
where $m$ is the mass of the target hadron (typically proton).
The reason we study the evolution of the operator (\ref{operator}) with non-convoluted indices $i$ and $j$ is that, as we shall see below, the rapidity evolution mixes functions $\cald$ and $\calh$. It should be also noted that our final equation for the evolution of the operator
(\ref{operator}) is applicable for polarized targets as well.

In the spirit of rapidity factorization, in order to find the evolution of TMD 
\begin{equation}
\langle\!\langle p|\calf^a_i(x_\ast, x_\perp)[x_\perp,y_\perp]_{-\infty}^{ab}\calf^b_j(y_\ast, y_\perp)|p\rangle\!\rangle ^\eta
\label{TMD}
\end{equation}
with respect to rapidity cutoff $\eta$ 
(see Eq. (\ref{cutoff})) one should integrate in the matrix element  (\ref{TMD}) over gluons and quarks with rapidities $\eta>Y>\eta'$ and temporarily ``freeze'' fields with $Y<\eta'$ to be integrated over later. (For a review, see Refs. \cite{mobzor,nlolecture}.) In this case, we obtain functional integral of Eq. (\ref{funtegral}) type over fields with $\eta>Y>\eta'$ in the ``external'' fields with $Y<\eta'$. 
In terms of Sudakov variables we integrate over gluons with $\alpha$ between $\sigma=e^\eta$ and $\sigma'=e^{\eta'}$ and, in the leading order, only
the diagrams with gluon emissions are relevant - the quark diagrams will enter as loops at the next-to-leading (NLO) level.

\begin{figure}[htb]
\begin{center}
\includegraphics[width=66mm]{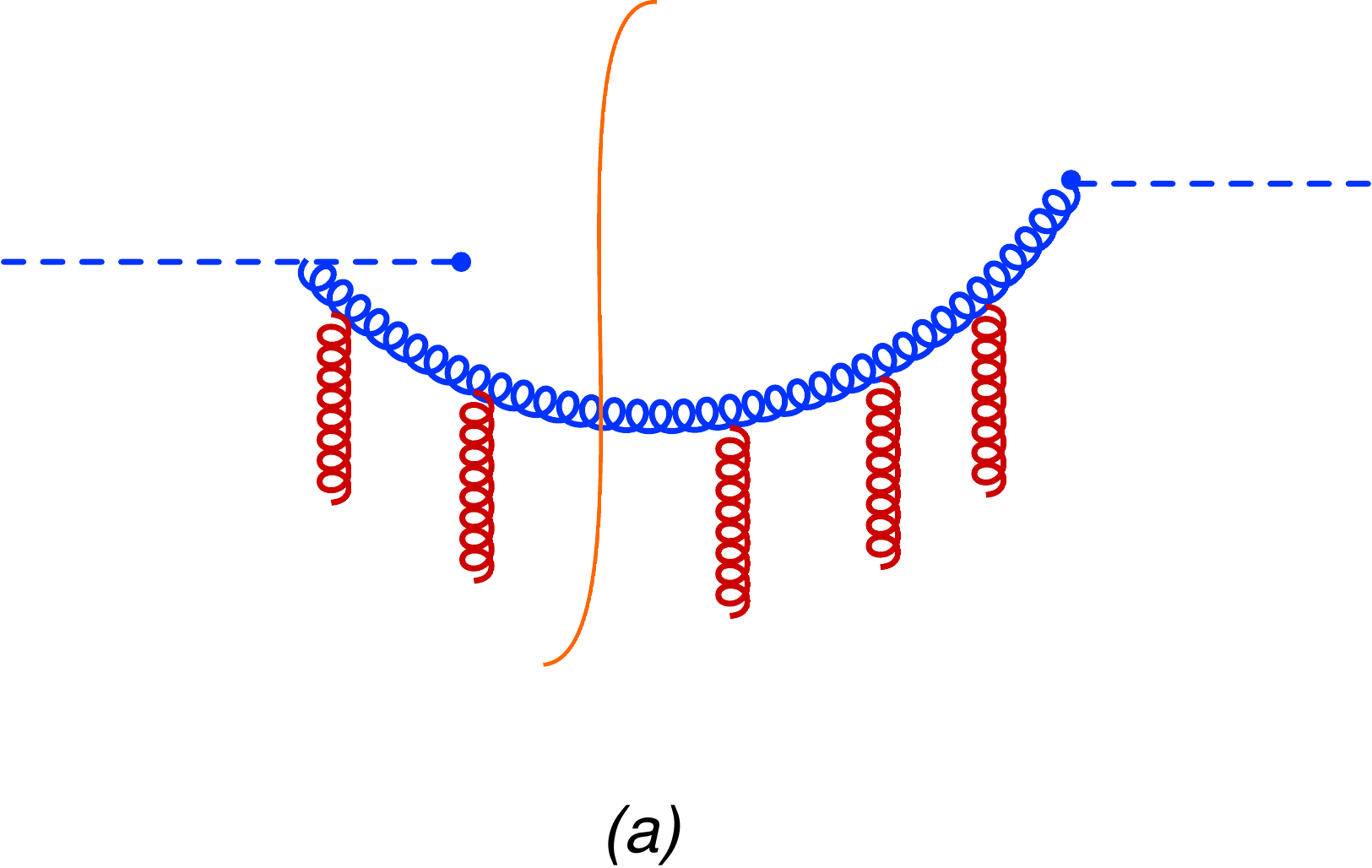}
\hspace{12mm}
\includegraphics[width=66mm]{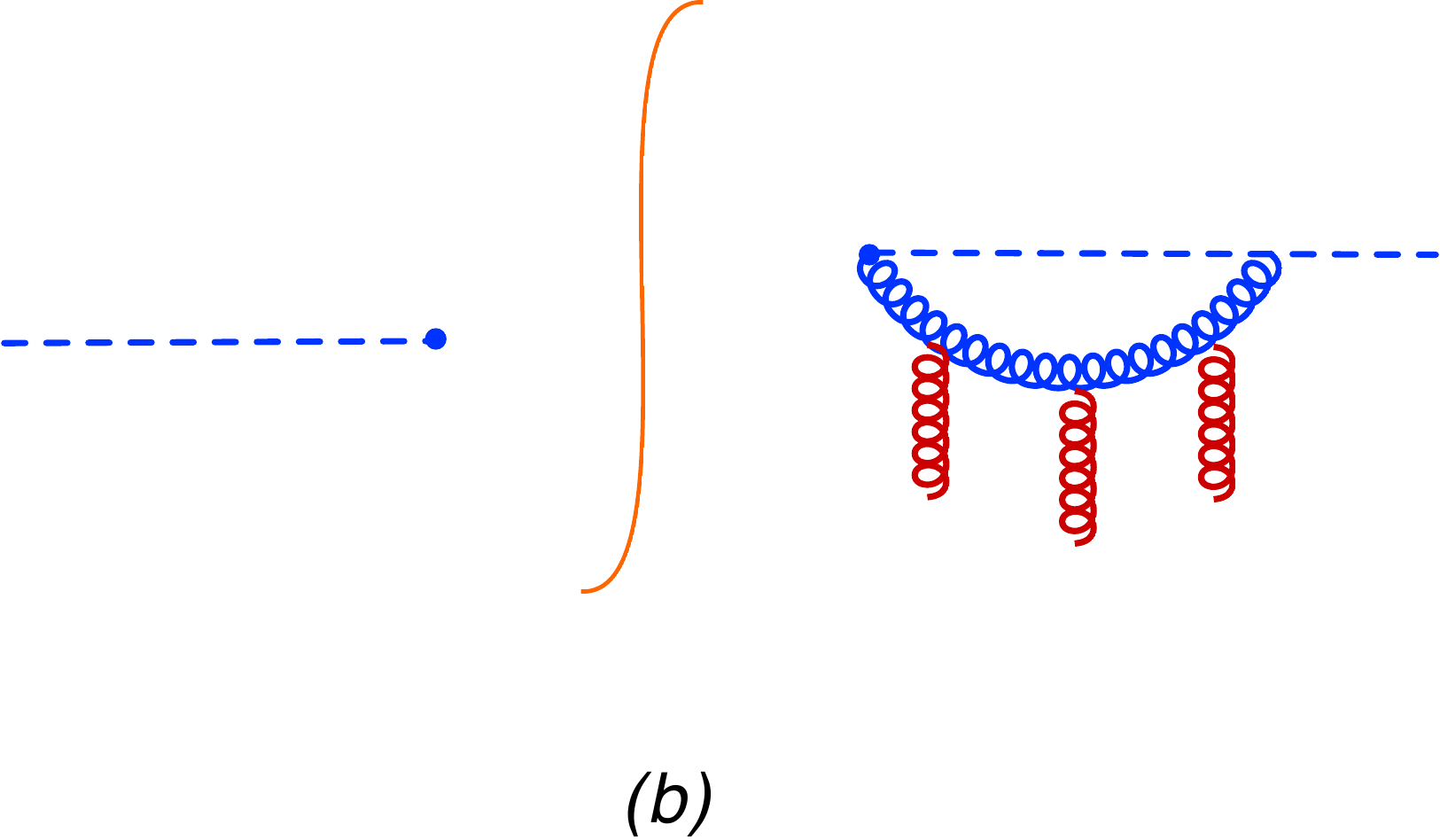}
\end{center}
\caption{Typical diagrams for production (a) and virtual (b) contributions to the evolution kernel. The dashed lines denote gauge links.\label{fig:2}}
\end{figure}

To calculate diagrams, one needs to return to a double functional integral representation of gluon TMD (\ref{TMD}):
\begin{eqnarray}
&&\hspace{-1mm}
\langle p|\calf^a_i(x_\ast, x_\perp)[x_\perp,y_\perp]_{-\infty}^{ab}\calf^b_j(y_\ast, y_\perp)|p'\rangle^\eta
\nonumber\\
&&\hspace{-1mm}
=~\int^{\tilA(\infty)=A(\infty)}\! D\tilA D\tilde{\bar\psi}D\tilde{\psi}DA D\bsi D\psi ~
\Psi^*_{p}(\tilA,\tipsi)|_{t_i=-\infty}~e^{-iS_{\rm QCD}(\tilA,\tipsi)}
\nonumber\\
&&\hspace{-1mm}
\ticalf^a_i(x_\ast, x_\perp)[x_\perp,y_\perp]_{-\infty}^{ab}
~e^{iS_{\rm QCD}(A,\psi)}
\calf^b_j(y_\ast, y_\perp)\Psi_{p'}(A,\psi)|_{t_i=-\infty}
\label{3.2}
\end{eqnarray}
Now, in accordance with general background-field formalism we separate the gluon field into 
the ``classical'' background part with $Y<\eta'$ and
``quantum'' part with $\eta>Y>\eta'$ and integrate over quantum fields.
In the leading order there are 
two types of diagrams: with and without gluon production, see Fig. \ref{fig:2}  (we assume that there are no gluons with
 $\eta>Y>\eta'$ in the proton wave function).

\subsection{Production part of the LO kernel\label{sec2}}

The first-order term in the expansion of the operator $F_{\bu i}^m(y_\ast,y_\perp)[y_\ast,-\infty]_y^{ma}$ in quantum fields has the form
\begin{eqnarray}
&&\hspace{-11mm}
F_{\bu i}^m(y_\ast,y_\perp)[y_\ast,-\infty]_y^{ma}~\stackrel{\rm 1st}{=}~
{s\over 2}{\partial\over\partial y_\ast}A_i^{m{\rm q}}(y_\ast,y_\perp)[y_\ast,-\infty]_y^{ma}
\label{3.3}\\
&&\hspace{-11mm}
-~\partial_iA_\bu^{m{\rm q}}(y_\ast,y_\perp)[y_\ast,-\infty]_y^{ma}
+i\!\int_{-\infty}^{y_\ast}\! d{2\over s}z'_\ast ~F_{\bu i}^m(y_\ast,y_\perp)([y_\ast,z'_\ast]_yA_\bu^{\rm q}(z'_\ast,y_\perp)[z'_\ast,-\infty]_y)^{ma}
\nonumber
\end{eqnarray}
(to save space, we omit the label ${}^{\rm cl}$ from classical fields).  As it was proved in Ref. \cite{gTMD1}, to find the evolution kernel
in the leading order in $\alpha_s$ it is sufficient to consider the classical background field of the form
\begin{equation}
A^{\rm cl}_\mu~=~{2\over s}p_{2\mu}A_\bu(x_\ast,x_\perp)
\label{klfild}
\end{equation}
where the absence of $x_\bu$ in the argument corresponds to $\alpha=0$. 

 Using the gluon propagator (\ref{glupromp}) from Sect. \ref{app:gluprop} we obtain the result for the diagram in Fig. \ref{fig:2}a in the form
\begin{eqnarray}
&&\hspace{-1mm}
\langle \ticalf_i^a(x_\ast,x_\perp)\calf^a_j(y_\ast,y_\perp)\rangle^{\eta>Y>\eta'} ~
\nonumber\\
&&\hspace{-1mm}
=~-{1\over 4}\!\int_{\sigma'}^\sigma\!{\dhd\alpha\over 2\alpha^3}
{\rm Tr}[-\infty,x_\ast]_x\Big[
(x_\perp|(p_\perp^2g_{ik}+2p_ip_k)e^{-i{p_\perp^2\over\alpha s}x_\ast}\ticalo_\alpha(p_\perp,x_\ast,\infty)
\label{3.5}\\
&&\hspace{22mm}
+~
{4\over s}\!\int^{x_\ast}_{-\infty} \! dx'_\ast \tilF_{\bu i}(x_\ast,x_\perp)
[x_\ast,x'_\ast]_x(x_\perp|p_ke^{-i{p_\perp^2\over\alpha s}x'_\ast}\ticalo_\alpha(p_\perp,x'_\ast,\infty)\Big]
\nonumber\\
&&\hspace{-1mm}
\times~\Big[\calo_\alpha(\infty,y_\ast,p_\perp)e^{i{p_\perp^2\over\alpha s}y_\ast}(p_\perp^2\delta_j^k+2p_jp^k)|y_\perp)
\nonumber\\
&&\hspace{22mm}
+~{4\over s}\calo_\alpha(\infty,y'_\ast,p_\perp)\!\int^{y_\ast}_{-\infty}\! dy'_\ast~e^{i{p_\perp^2\over\alpha s}y'_\ast}p^k|y_\perp)
[y'_\ast,y_\ast]_yF_{\bu j}(y_\ast,y_\perp)\Big][y_\ast,-\infty]_y
\nonumber
\end{eqnarray}
where $\langle O\rangle$ denotes the expectation value of operator $O$ in the external field.
Note that in this paper we perform calculations of diagrams in the background field (\ref{klfild}) in the light-like gauge
\begin{equation}
p_2^\mu A_\mu(x)~=~0
\label{llgauge}
\end{equation}
We will make necessary comparisons with the background-Feynman gauge calculations of Ref. \cite{gTMD1} in Appendix \ref{app:compare}.

Let us consider now the remaining integral over ``classical''  fields with $Y<\eta'$. It has the form
\begin{eqnarray}
&&\hspace{-1mm}
-{1\over 4}\!\int_{\sigma'}^\sigma\!{\dhd\alpha\over 2\alpha^3}
\int^{\tilA(\infty)=A(\infty)}\! D\tilA D\tilde{\bar\psi}D\tilde{\psi}DA D\bsi D\psi ~
~e^{-iS_{\rm QCD}(\tilA,\tipsi)}
~e^{iS_{\rm QCD}(A,\psi)}
\nonumber\\
&&\hspace{-1mm}
\times~\Psi^\ast_p(\tilA,\tipsi)|_{t_i=-\infty}
{\rm Tr}[-\infty,x_\ast]_x\Big[
(x_\perp|(p_\perp^2g_{ik}+2p_ip_k)e^{-i{p_\perp^2\over\alpha s}x_\ast}\ticalo_\alpha(p_\perp,x_\ast,\infty)
\label{doublefun}\\
&&\hspace{11mm}
+~
{4\over s}\!\int^{x_\ast}_{-\infty} \! dx'_\ast \tilF_{\bu i}(x_\ast,x_\perp)
[x_\ast,x'_\ast]_x(x_\perp|p_ke^{-i{p_\perp^2\over\alpha s}x'_\ast}\ticalo_\alpha(p_\perp,x'_\ast,\infty)\Big]
\nonumber\\
&&\hspace{-1mm}
\times~\Big[\calo_\alpha(\infty,y_\ast,p_\perp)e^{i{p_\perp^2\over\alpha s}y_\ast}(p_\perp^2\delta_j^k+2p_jp^k)|y_\perp)
\nonumber\\
&&\hspace{-1mm}
+~{4\over s}\calo_\alpha(\infty,y'_\ast,p_\perp)\!\int^{y_\ast}_{-\infty} \! dy'_\ast~e^{i{p_\perp^2\over\alpha s}y'_\ast}p^k|y_\perp)
[y'_\ast,y_\ast]_yF_{\bu j}(y_\ast,y_\perp)\Big][y_\ast,-\infty]_y\Psi_{p'}(A,\psi)|_{t_i=-\infty}
\nonumber
\end{eqnarray}
where ${\rm Tr}(...)$ is the trace in the adjoint representation. As discussed in Appendix \ref{appcrsc}, the double functional 
integral (\ref{doublefun}) represents the matrix element 
\begin{eqnarray}
&&\hspace{-1mm}
-{1\over 4}\!\int_{\sigma'}^\sigma\!{\dhd\alpha\over 2\alpha^3}
\langle p|{\rm Tr}[-\infty,x_\ast]_x\Big[
(x_\perp|(p_\perp^2g_{ik}+2p_ip_k)e^{-i{p_\perp^2\over\alpha s}x_\ast}\calo_\alpha(p_\perp,x_\ast,\infty)
\label{3.9}\\
&&\hspace{22mm}
+~
{4\over s}\!\int^{x_\ast}_{-\infty}\! dx'_\ast F_{\bu i}(x_\ast,x_\perp)
[x_\ast,x'_\ast]_x(x_\perp|p_ke^{-i{p_\perp^2\over\alpha s}x'_\ast}\calo_\alpha(p_\perp,x'_\ast,\infty)\Big]
\nonumber\\
&&\hspace{-1mm}
\times~\Big[\calo_\alpha(\infty,y_\ast,p_\perp)e^{i{p_\perp^2\over\alpha s}y_\ast}(p_\perp^2\delta_j^k+2p_jp^k)|y_\perp)
\nonumber\\
&&\hspace{22mm}
+~{4\over s}\calo_\alpha(\infty,y'_\ast,p_\perp)\!\int^{y_\ast}_{-\infty}\! dy'_\ast~e^{i{p_\perp^2\over\alpha s}y'_\ast}p^k|y_\perp)
[y'_\ast,y_\ast]_yF_{\bu j}(y_\ast,y_\perp)\Big][y_\ast,-\infty]_y|p'\rangle
\nonumber
\end{eqnarray}
As we mentioned above, all operators in the r.h.s. of Eq. (\ref{3.9}) commute since they are separated either by 
space-like or by light-like distance.  In addition, from Eq. (\ref{grpr}) we see that
\begin{equation}
\calo_\alpha(p_\perp,x_\ast,\infty)\calo_\alpha(\infty,y_\ast,p_\perp)~=~\calo_\alpha(x_\ast,y_\ast)
~=~\calo_\alpha(p_\perp,x_\ast,-\infty)\calo_\alpha(-\infty,y_\ast,p_\perp)
\label{grouplaw}
\end{equation}
Substituting Eq. (\ref{grouplaw}) in Eq. (\ref{3.5}) we get
\begin{eqnarray}
&&\hspace{-1mm}
\langle\!\langle p|\calf^a_i(x_\ast, x_\perp)\calf^a_j(y_\ast, y_\perp)|p\rangle\!\rangle ^\eta
\label{master1}\\
&&\hspace{-1mm}
=~
-{1\over 4}\!\int_{\sigma'}^\sigma\!{\dhd\alpha\over 2\alpha^3}
\langle\!\langle p|{\rm Tr}[-\infty,x_\ast]_x\Big[
(x_\perp|(p_\perp^2g_{ik}+2p_ip_k)e^{-i{p_\perp^2\over\alpha s}x_\ast}\calo_\alpha(p_\perp,x_\ast,-\infty)
\nonumber\\
&&\hspace{22mm}
+~
{4\over s}\!\int^{x_\ast}_{-\infty} \! dx'_\ast F_{\bu i}(x_\ast,x_\perp)
[x_\ast,x'_\ast]_x(x_\perp|p_ke^{-i{p_\perp^2\over\alpha s}x'_\ast}\calo_\alpha(p_\perp,x'_\ast,-\infty)\Big]
\nonumber\\
&&\hspace{-1mm}
\times~\Big[\calo_\alpha(-\infty,y_\ast,p_\perp)e^{i{p_\perp^2\over\alpha s}y_\ast}(p_\perp^2\delta_j^k+2p_jp^k)|y_\perp)
\nonumber\\
&&\hspace{22mm}
+~{4\over s}\calo_\alpha(-\infty,y'_\ast,p_\perp)\!\int^{y_\ast}_{-\infty}\! dy'_\ast~e^{i{p_\perp^2\over\alpha s}y'_\ast}p^k|y_\perp)
[y'_\ast,y_\ast]_yF_{\bu j}(y_\ast,y_\perp)\Big][y_\ast,-\infty]_y|p\rangle\!\rangle 
\nonumber
\end{eqnarray}
At this point we compare (\ref{master1}) to the evolution equation for \\
$\langle\!\langle p| F^m_{\bu i}(x_\ast,x_\perp)[x_\ast,\infty]^{ma}[\infty,y_\ast]^{an}F^n_{\bu j}(y_\ast,y_\perp)|p\rangle\!\rangle$.
Repeating steps which lead us from Eq. (\ref{3.5}) to Eq. (\ref{master1}) we obtain
\begin{eqnarray}
&&\hspace{-1mm}
\langle\!\langle p| F^m_{\bu i}(x_\ast,x_\perp)[x_\ast,\infty]^{ma}[\infty,y_\ast]^{an}F^n_{\bu j}(y_\ast,y_\perp)
|p\rangle\!\rangle^\eta
\nonumber\\
&&\hspace{-1mm}
=~
-{1\over 4}\!\int_{\sigma'}^\sigma\!{\dhd\alpha\over 2\alpha^3}
\langle\!\langle p|{\rm Tr}[\infty,x_\ast]_x\Big[
(x_\perp|(p_\perp^2g_{ik}+2p_ip_k)e^{-i{p_\perp^2\over\alpha s}x_\ast}\calo_\alpha(p_\perp,x_\ast,\infty)
\nonumber\\
&&\hspace{22mm}
-~
{4\over s}\!\int_{x_\ast}^{\infty} \! dx'_\ast F_{\bu i}(x_\ast,x_\perp)
[x_\ast,x'_\ast]_x(x_\perp|p_ke^{-i{p_\perp^2\over\alpha s}x'_\ast}\calo_\alpha(p_\perp,x'_\ast,\infty)\Big]
\nonumber\\
&&\hspace{-1mm}
\times~\Big[\calo_\alpha(\infty,y_\ast,p_\perp)e^{i{p_\perp^2\over\alpha s}y_\ast}(p_\perp^2\delta_j^k+2p_jp^k)|y_\perp)
\label{masterplus}\\
&&\hspace{22mm}
-~{4\over s}\calo_\alpha(\infty,y'_\ast,p_\perp)\!\int_{y_\ast}^\infty \! dy'_\ast~e^{i{p_\perp^2\over\alpha s}y'_\ast}p^k|y_\perp)
[y'_\ast,y_\ast]_yF_{\bu j}(y_\ast,y_\perp)\Big][y_\ast,\infty]_y|p\rangle\!\rangle
\nonumber
\end{eqnarray}
We see that the production part of the evolution equation (\ref{master1}) can be obtained from Eq. (\ref{masterplus}) by formally replacing $+\infty$ 
by $-\infty$ everywhere.  Consequently,  the final expression for the production part of the evolution equation for the matrix 
element (\ref{TMD}) can be obtained from Eq.  (4.28)  from Ref. \cite{gTMD1}
by replacement $\infty\leftrightarrow -\infty$. 
\footnote{In the Appendix \ref{app:compare} we show that the Eq. (\ref{masterplus}), obtained in the  light-like gauge,
agrees with the calculations in  Ref. \cite{gTMD1} performed in the background-Feynman gauge.}

\subsection{Virtual part of the evolution kernel}

The virtual part of the kernel comes from the diagrams of the Fig. \ref{fig:2}b type. The  second-order term in the expansion of the 
operator $F_{\bu i}^m(y_\ast,y_\perp)[y_\ast,-\infty]_y^{ma}$ in quantum fields has the form (cf. Eq. (\ref{3.3}))
\begin{eqnarray}
&&\hspace{-1mm}
F_{\bu i}^m(y_\ast,y_\perp)[y_\ast,-\infty]_y^{ma}~\stackrel{\rm 2nd}{=}~
\nonumber\\
&&\hspace{-1mm}
=~i\!\int_{-\infty}^{y_\ast}\! \!d{2\over s}z'_\ast ~(D_\bu A_i^{m q}-\partial_iA_\bu^{m q})(y_\ast)[y_\ast,z'_\ast]A^q_\bu(z'_\ast)[z'_\ast,-\infty]^{ma}
+f^{mcd}A^{c q}_\bu A^{d q}_i[y_\ast,-\infty]^{ma}
\nonumber\\
&&\hspace{-1mm}
-~\!\int_{-\infty}^{y_\ast}d{2\over s}z'_\ast\!\int^{z'_\ast}_{-\infty}\! d{2\over s}z''_\ast ~
F_{\bu i}^m(y_\ast,y_\perp)\big([y_\ast,z'_\ast]A^q_\bu(z'_\ast)[z'_\ast,z''_\ast]A^q_\bu(z''_\ast)[z''_\ast, -\infty]\big)^{ma}
\label{4.1}
\end{eqnarray}
Using gluon propagator (\ref{glupropaxi}) we get
\begin{eqnarray}
&&\hspace{-2mm}
F_{\bu i}^m(y_\ast,y_\perp)[y_\ast,-\infty]_y^{ma}~\stackrel{\rm 2nd}{=}~{i\over s}\!\int^{y_\ast}_{-\infty}\! dy'_\ast~
{\rm Tr}T^a[-\infty,y_\ast](y_\perp,y_\ast|(p_\perp^2\delta_i^j+2p_ip^j){1\over\alpha^2 P^2}p_j
\label{4.2}\\
&&\hspace{-2mm}
+p_i{1\over\alpha^2 P^2}p_\perp^2
|y_\perp,y'_\ast)
[y'_\ast,-\infty]+~{4i\over s^2}\!\int_{-\infty}^{y_\ast}dy'_\ast\!\int^{y'_\ast}_{-\infty}\! dy''_\ast ~
{\rm Tr} T^a[-\infty,y_\ast]F_{\bu i}(y_\ast,y_\perp)[y_\ast,y'_\ast]_y
\nonumber\\
&&\hspace{45mm}
\times(y_\perp,y'_\ast|p^j{1\over\alpha^2P^2}p_j-{1\over\alpha^2}|y_\perp,y''_\ast)
[y''_\ast,-\infty]_y
\nonumber\\
&&\hspace{-2mm}
=~{1\over s}\int_0^\infty{\dhd\alpha\over 2\alpha^3}\!\int^{y_\ast}_{-\infty}\! dy'_\ast~   
{\rm Tr}T^a[-\infty,y_\ast](y_\perp|e^{-i{p_\perp^2\over\alpha s}y_\ast}\big\{
(p_\perp^2\delta_i^j+2p_ip^j)\calo_\alpha(y_\ast,y'_\ast) p_j
\nonumber\\
&&\hspace{-2mm}
+~p_i\calo_\alpha(y_\ast,y'_\ast)p_\perp^2\big\}e^{i{p_\perp^2\over\alpha s}y'_\ast}|y_\perp)
[y'_\ast,-\infty]+~{4i\over s^2}\!\int_{-\infty}^{y_\ast}dy'_\ast\!\int^{y'_\ast}_{-\infty}\! dy''_\ast ~
{\rm Tr} T^a[-\infty,y_\ast]F_{\bu i}(y_\ast,y_\perp)
\nonumber\\
&&\hspace{45mm}
\times [y_\ast,y'_\ast]_y(y_\perp,y'_\ast|p^j{1\over\alpha^2P^2}p_j-{1\over\alpha^2}|y_\perp,y''_\ast)
[y''_\ast,-\infty]_y
\nonumber
\end{eqnarray}
Let us start with the last term in the r.h.s. of the above equation.  We will prove that
\begin{equation}
(y_\perp,y'_\ast|p^j{1\over\alpha^2P^2}p_j-{1\over\alpha^2}|y_\perp,y''_\ast)
~=~[y'_\ast,y''_\ast]_y(y_\perp,y'_\ast|p^j{1\over\alpha^2p^2}p_j-{1\over\alpha^2}|y_\perp,y''_\ast)
\label{4.3}
\end{equation}
in our approximation. Indeed, in the ``light-cone'' case (when the characteristic transverse momenta
of background field $l_\perp$ are much smaller than the momenta of the ``quantum'' fields $p_\perp$) 
it is evident since 
\begin{equation}
(y'_\ast,y_\perp|{1\over P^2}|y''_\ast,y_\perp)~=~(y'_\ast,y_\perp|{1\over p^2}|y''_\ast,y_\perp)
[y'_\ast,y''_\ast]_y~+~O(F_{\bu j})
\label{4.4}
\end{equation}
and terms $\sim O(F_{\bu j})$ exceed our accuracy. (The second term in the l.h.s. of Eq. (\ref{4.3})  is proportional
to $\delta(y'_\ast-y''_\ast)$ and $[y'_\ast,y''_\ast]_y=1$ is introduced for convenience.)

In the ``shock-wave'' case when $l_\perp\sim p_\perp$,
if the points $y'$ and $y''$ are outside of the shock wave, the formula is trivial ($y'$ and $y''$ can only be both to the right  of the shock wave since $y$ lies inside). If $y'$ or both of them are inside the shock wave, one can again use the light-cone expansion
(see the discussion in Ref. \cite{gTMD1}) and get the result (\ref{4.4}). Thus, in both cases we can use Eq. (\ref{4.3}) so
\begin{eqnarray}
&&\hspace{-2mm}
{4i\over s^2}~
N_c\calf^a_i(y_\ast, y_\perp)\!\int_{-\infty}^{y_\ast}dy'_\ast\!\int^{y'_\ast}_{-\infty}\! dy''_\ast 
(y_\perp,y'_\ast|p^j{1\over\alpha^2(p^2+i\epsilon)}p_j-{1\over\alpha^2}|y_\perp,y''_\ast)
\label{4.5}\\
&&\hspace{-2mm}
=~-2i
N_c\calf^a_i(y_\ast, y_\perp)\!\int_{-\infty}^{y_\ast}dy'_\ast\!\int^{y'_\ast}_{-\infty}\! dy''_\ast 
\!\int\!\dhd\alpha\dhd\beta ~e^{-i\beta(y'_\ast-y''_\ast)}
(y_\perp|{\beta\over\alpha(\alpha\beta s-p_\perp^2+i\epsilon)}|y_\perp)
\nonumber
\end{eqnarray}
where we used formula ${\rm Tr} T^a[-\infty,y_\ast]F_{\bu i}(y_\ast,y_\perp)[y_\ast,-\infty]_y~=~N_c\calf^a_i(y_\ast, y_\perp)$. It is convenient
to change $\alpha\leftrightarrow-\alpha$ and $\beta\leftrightarrow-\beta$ (which is equivalent to changing $y'_\ast\leftrightarrow y''_\ast$)
and get
\begin{eqnarray}
&&\hspace{-2mm}
-iN_c\calf^a_i(y_\ast, y_\perp)\!\int_{-\infty}^{y_\ast}dy'_\ast dy''_\ast 
\!\int\!\dhd\alpha\dhd\beta ~e^{-i\beta(y'_\ast-y''_\ast)}
(y_\perp|{\beta\over\alpha(\alpha\beta s-p_\perp^2+i\epsilon)}|y_\perp)
\nonumber\\
&&\hspace{-2mm}
=~-iN_c\calf^a_i(y_\ast, y_\perp)
\!\int\! {\dhd\alpha\over\alpha}\dhd\beta ~{\rm V.p.}{1\over\beta}
(y_\perp|{1\over\alpha\beta s-p_\perp^2+i\epsilon}|y_\perp)
\end{eqnarray}
where V.p. means principle value: ${\rm V.p.}{1\over x}\equiv \half\big({1\over x-i\epsilon}+{1\over x+i\epsilon}\big)$. Thus, we obtain the result for the last term in the r.h.s. of Eq. (\ref{4.2}) in the form
\begin{eqnarray}
&&\hspace{-2mm}
{4i\over s^2}\!\int_{-\infty}^{y_\ast}dy'_\ast\!\int^{y'_\ast}_{-\infty}\! dy''_\ast ~
{\rm Tr} T^a[-\infty,y_\ast]F_{\bu i}(y_\ast)[y_\ast,y'_\ast]_y(y_\perp,y'_\ast|p^j{1\over\alpha^2P^2}p_j-{1\over\alpha^2}|y_\perp,y''_\ast)[y''_\ast,-\infty]_y
\nonumber\\
&&\hspace{25mm}
=~-{N_c\over 2}\calf^a_i(y_\ast, y_\perp)(y_\perp|{1\over p_\perp^2}|y_\perp)
\Big[\!\int_0^\infty\! {\dhd\alpha\over\alpha}-\!\int^0_{-\infty}\! {\dhd\alpha\over\alpha}\Big]
\nonumber\\
&&\hspace{25mm}
=~-N_c\calf^a_i(y_\ast, y_\perp)(y_\perp|{1\over p_\perp^2}|y_\perp)\!\int_0^\infty\! {\dhd\alpha \over\alpha}
\label{4.7}
\end{eqnarray}

Next we turn our attention to the first term in the r.h.s. of Eq. (\ref{4.2}) and start with the light-cone case $l_\perp\ll p_\perp$:
\begin{eqnarray}
&&\hspace{-1mm}
{1\over s}\int_0^\infty{\dhd\alpha\over 2\alpha^3}\!\int^{y_\ast}_{-\infty}\! dy'_\ast~   
{\rm Tr}T^a[-\infty,y_\ast](y_\perp|e^{-i{p_\perp^2\over\alpha s}y_\ast}\big\{
(p_\perp^2\delta_i^j+2p_ip^j)\calo_\alpha(y_\ast,y'_\ast) p_j
\nonumber\\
&&\hspace{22mm}
+~p_i\calo_\alpha(y_\ast,y'_\ast)p_\perp^2\big\}e^{i{p_\perp^2\over\alpha s}y'_\ast}|y_\perp)
[y'_\ast,-\infty]
\nonumber\\
&&\hspace{-1mm}
={1\over s}\int_0^\infty{\dhd\alpha\over 2\alpha^3}\!\int^{y_\ast}_{-\infty}\! dy'_\ast~   
{\rm Tr}T^a[-\infty,y_\ast](y_\perp|e^{-i{p_\perp^2\over\alpha s}(y-y')_\ast}
(p_\perp^2\delta_i^j+2p_ip^j)\calo_\alpha^{y'_\ast}(y_\ast,y'_\ast) p_j
\nonumber\\
&&\hspace{22mm}
+~p_i\calo_\alpha^{y_\ast}(y_\ast,y'_\ast)p_\perp^2e^{-i{p_\perp^2\over\alpha s}(y-y')_\ast}|y_\perp)
[y'_\ast,-\infty]
\label{4.8}
\end{eqnarray}
where $\calo_\alpha^{y'_\ast}$ is defined in Eq. (\ref{oa}). The first term in the r.h.s.  of this equation yields
\begin{eqnarray}
&&\hspace{-2mm}
\int_0^\infty{\dhd\alpha\over 2\alpha^3s}\!\int^{y_\ast}_{-\infty}\! \!\!dy'_\ast~   
{\rm Tr}T^a(y_\perp|e^{-i{p_\perp^2\over\alpha s}(y-y')_\ast}
\Big((p_\perp^2g_{ij}+2p_ip_j)\!\int_{y'_\ast}^{y_\ast}\! d{2\over s}z_\ast [-\infty,z_\ast]F_{\bu j}(z_\ast)[z_\ast,-\infty]_y
\nonumber\\
&&\hspace{1mm}
+~{4ig\over\alpha s^2}p_\perp^2p_ip^j
\!\int_{y'_\ast}^{y_\ast}\!\!\!dz_\ast~(z-y')_\ast[-\infty,z_\ast]F_{\bu j}(z_\ast)[z_\ast,-\infty]_y\Big)|y_\perp)
\\
&&\hspace{-1mm}
=~-ig\int_0^\infty{\dhd\alpha\over \alpha^4s^3}\!\int^{y_\ast}_{-\infty}\! \!\!dy'_\ast~   
(y_\perp|p_\perp^4e^{-i{p_\perp^2\over\alpha s}(y-y')_\ast}|y_\perp)
\nonumber\\
&&\hspace{65mm}
\times\int_{y'_\ast}^{y_\ast}\!\!\!dz_\ast~(z-y')_\ast{\rm Tr}T^a[-\infty,z_\ast]F_{\bu i}(z_\ast)[z_\ast,-\infty]_y
\nonumber
\end{eqnarray}
so one obtains
\begin{eqnarray}
&&\hspace{-1mm}
\langle(D_\bu A^m_i-\partial_iA^m_\bu)(y_\ast,y_\perp)[y_\ast,-\infty]^{ma}\rangle~=~{1\over s}\int_0^\infty{\dhd\alpha\over 2\alpha^3}\!\int^{y_\ast}_{-\infty}\! dy'_\ast~   
{\rm Tr}T^a[-\infty,y_\ast]
\nonumber\\
&&\hspace{-1mm}
\times(y_\perp|e^{-i\frac{p^2_\perp}{\alpha s}y_\ast}
(p_\perp^2\delta_i^j+2p_ip^j)\calo_\alpha(y_\ast,y'_\ast) p_je^{i\frac{p^2_\perp}{\alpha s}y'_\ast}|y_\perp)[y'_\ast, -\infty]
\label{4.10}\\
&&\hspace{22mm}
~=~igN_c\!\int_0^\infty{\dhd\alpha\over \alpha^2 s}\!\int^{y_\ast}_{-\infty}\! \!\!dz_\ast~   
(y_\perp|e^{-i{p_\perp^2\over\alpha s}(y-z)_\ast}|y_\perp)
F^m_{\bu i}(z_\ast)[z_\ast,-\infty]_y^{ma}
\nonumber
\end{eqnarray}
As to the second term in the r.h.s of Eq. (\ref{4.8}), it vanishes
\begin{eqnarray}
&&\hspace{-1mm}
\int_0^\infty{\dhd\alpha\over 2\alpha^3}\!\int^{y_\ast}_{-\infty}\! \!\!dy'_\ast~   
{\rm Tr}T^a[-\infty,y_\ast](y_\perp|
p_i\calo_\alpha^{y_\ast}(y_\ast,y'_\ast)p_\perp^2e^{-i{p_\perp^2\over\alpha s}(y-y')_\ast}|y_\perp)
[y'_\ast,-\infty]
\nonumber\\
&&\hspace{-1mm}
=~\int_0^\infty{\dhd\alpha\over 2\alpha^3}\!\int^{y_\ast}_{-\infty}\! \!\!dy'_\ast~   
{\rm Tr}T^a[-\infty,y_\ast](y_\perp|
\!\int_{y'_\ast}^{y_\ast}\! d{2\over s}z_\ast [y_\ast,z_\ast]F_{\bu i}(y_\perp,z_\ast)[z_\ast,y'_\ast]
p_\perp^2e^{-i{p_\perp^2\over\alpha s}(y-y')_\ast}
\nonumber\\
&&\hspace{-1mm}
-~{4ig\over\alpha s^2}\!\int_{y'_\ast}^{y_\ast}\!\!\!dz_\ast~(z-y)_\ast[y_\ast,z_\ast]F_{\bu j}(z_\ast)[z_\ast,y'_\ast]
p_\perp^2p_ip^je^{-i{p_\perp^2\over\alpha s}(y-y')_\ast}|y_\perp)[y'_\ast,-\infty]
\nonumber\\
&&\hspace{-1mm}
=~-ig\int_0^\infty{\dhd\alpha\over \alpha^2}\!\int^{y_\ast}_{-\infty}\! \!\!dz_\ast~   
(y_\perp|\Big(1-{ip_\perp^2(y-z)_\ast\over \alpha s}\Big)e^{-i{p_\perp^2\over\alpha s}(y-z)_\ast}|y_\perp)
\nonumber\\
&&\hspace{75mm}
\times{\rm Tr}T^a[-\infty,z_\ast]F_{\bu i}(z_\ast)[z_\ast,-\infty]_y~=~0
\nonumber
\end{eqnarray}
so
\begin{eqnarray}
&&\hspace{-1mm}
f^{mcd}A^{c q}_\bu A^{d q}_i[y_\ast,-\infty]^{ma}~=~0
\end{eqnarray}
in our approximation. Thus, the first term in r.h.s. of Eq. (\ref{4.2}) in the light-cone case has the form
\begin{eqnarray}
&&\hspace{-2mm}
\langle(D_\bu A^m_i-\partial_iA^m_\bu+gf^{mcd}A^{c q}_\bu A^{d q}_i)(y_\ast,y_\perp)
[y_\ast,-\infty]_y^{ma}\rangle
\nonumber\\
&&\hspace{-1mm}
=~igN_c\!\int_0^\infty{\dhd\alpha\over \alpha^2 s}\!\int^{y_\ast}_{-\infty}\! \!\!dz_\ast~   
(y_\perp|e^{-i{p_\perp^2\over\alpha s}(y-z)_\ast}|y_\perp)
F^m_{\bu i}(z_\ast)[z_\ast,-\infty]_y^{ma}
\label{4.12}
\end{eqnarray}
Let us now consider the shock-wave case. It is convenient to start with the representation of this 
term by the second line in Eq. (\ref{4.2}) 
\begin{eqnarray}
&&\hspace{-2mm}
\langle(D_\bu A^m_i-\partial_iA^m_\bu+gf^{mcd}A^{c q}_\bu A^{d q}_i)(y_\ast,y_\perp)
[y_\ast,-\infty]_y^{ma}\rangle~=~
\label{4.11}\\
&&\hspace{-2mm}
=~{i\over s}\!\int^{y_\ast}_{-\infty}\! dy'_\ast~
{\rm Tr}T^a[-\infty,y_\ast](y_\perp,y_\ast|(p_\perp^2\delta_i^j+2p_ip^j){1\over\alpha^2 P^2}p_j+p_i{1\over\alpha^2 P^2}p_\perp^2
|y_\perp,y'_\ast)[y'_\ast,-\infty]
\nonumber\\
&&\hspace{-2mm}
=~-{i\over s}\!\int^{y_\ast}_{-\infty}\! dy'_\ast~
{\rm Tr}T^a[-\infty,y_\ast](y_\perp,y_\ast|\big[p^j, [p_j,{1\over \alpha^2P^2}]\big]p_i
\nonumber\\
&&\hspace{45mm}
-2\big[p_i,[p^j,{1\over \alpha^2P^2}]\big]p_j+\big[p_i,{1\over \alpha^2P^2}\big]p_\perp^2
|y_\perp,y'_\ast)[y'_\ast,-\infty]
\nonumber
\end{eqnarray}
Using Eq. (\ref{pa1}) for Feynman propagator one obtains
\begin{eqnarray}
&&\hspace{-2mm}
\langle(D_\bu A^m_i-\partial_iA^m_\bu+gf^{mcd}A^{c q}_\bu A^{d q}_i)(y_\ast,y_\perp)
[y_\ast,-\infty]_y^{ma}\rangle~=~
-{1\over 2s}\!\int_0^\infty\!{\dhd\alpha\over\alpha^3}\!\int^{y_\ast}_{-\infty}\! dy'_\ast
\nonumber\\
&&\hspace{-2mm}
\times~~
{\rm Tr}T^a[-\infty,y_\ast](y_\perp|e^{-i{p_\perp^2\over\alpha s}y_\ast}\big\{(\delta_i^j\partial_\perp^2+2\partial_i\partial^j)
\calo(y_\ast,y'_\ast)p_j
\nonumber\\
&&\hspace{55mm}
+i\partial_i\calo(y_\ast,y'_\ast)p_\perp^2
\big\}e^{i{p_\perp^2\over\alpha s}y'_\ast}|y_\perp)[y'_\ast,-\infty]
\label{4.14}
\end{eqnarray}

If the point $y_\ast$ is outside the shock wave this gives 
\begin{eqnarray}
&&\hspace{-2mm}
-{\theta(y_\ast)\over 2s}\!\int_0^\infty\!{\dhd\alpha\over\alpha^3}\!\int^{y_\ast}_{-\infty}\! dy'_\ast
{\rm Tr}T^aU_y^\dagger(y_\perp|e^{-i{p_\perp^2\over\alpha s}y_\ast}\big\{\delta_i^j\partial_\perp^2Up_j+2\partial_i\partial^jU
p_j+i\partial_iUp_\perp^2\big\}e^{i{p_\perp^2\over\alpha s}y'_\ast}|y_\perp)
\nonumber\\
&&\hspace{-1mm}
=~i\theta(y_\ast)\!\int_0^\infty\!{\dhd\alpha\over 2\alpha^2}
{\rm Tr}T^aU_y^\dagger(y_\perp|e^{-i{p_\perp^2\over\alpha s}y_\ast}\big\{(\delta_i^j\partial_\perp^2+2\partial_i\partial^j)U
{p_j\over p_\perp^2}+i\partial_iU\big\}|y_\perp)
\label{4.15}
\end{eqnarray}
If the point $y$ is inside the shock wave we can again use the light-cone expansion and get Eq. (\ref{4.12}). It is easy to see that 
in both cases we can approximate the first term in Eq. (\ref{4.2})  by
\begin{eqnarray}
&&\hspace{-2mm}
\langle(D_\bu A^m_i-\partial_iA^m_\bu+gf^{mcd}A^{c q}_\bu A^{d q}_i)(y_\ast,y_\perp)
[y_\ast,-\infty]_y^{ma}\rangle
\nonumber\\
&&\hspace{-1mm}
=~i\theta(y_\ast)\!\int_0^\infty\!{\dhd\alpha\over 2\alpha^2}
{\rm Tr}T^aU_y^\dagger(y_\perp|e^{-i{p_\perp^2\over\alpha s}y_\ast}(\delta_i^j\partial_\perp^2U+2\partial_i\partial^jU)
{p_j\over p_\perp^2}|y_\perp)
\nonumber\\
&&\hspace{-1mm}
+~igN_c\!\int_0^\infty{\dhd\alpha\over \alpha^2 s}\!\int^{y_\ast}_{-\infty}\! \!\!dz_\ast~   
(y_\perp|e^{-i{p_\perp^2\over\alpha s}(y-z)_\ast}|y_\perp)
F^m_{\bu i}(z_\ast)[z_\ast,-\infty]_y^{ma}
\end{eqnarray}
with our accuracy. Adding the contribution (\ref{4.7}) of the second term in r.h.s. of Eq. (\ref{4.2}) we finally obtain the 
second-order virtual correction in the form
\begin{eqnarray}
&&\hspace{-1mm}
F_{\bu i}^m(y_\ast,y_\perp)[y_\ast,-\infty]_y^{ma}~\stackrel{\rm 2nd}{=}~
-~N_cF_{\bu i}^m(y_\ast,y_\perp)[y_\ast,-\infty]_y^{ma}
(y_\perp|{1\over p_\perp^2}|y_\perp)\!\int_{\sigma'}^\sigma\!{\dhd\alpha \over\alpha}
\nonumber\\
&&\hspace{11mm}
+~igN_c\!\int_{\sigma'}^\sigma\!{\dhd\alpha\over \alpha^2 s}\!\int^{y_\ast}_{-\infty}\! \!\!dy'_\ast~   
(y_\perp|e^{-i{p_\perp^2\over\alpha s}(y-y')_\ast}|y_\perp)
F^m_{\bu i}(y'_\ast,y_\perp)[y'_\ast,-\infty]_y^{ma}
\nonumber\\
&&\hspace{11mm}
+~i\theta(y_\ast)\!\int_{\sigma'}^\sigma\!{\dhd\alpha\over 2\alpha^2}
{\rm Tr}T^aU_y^\dagger(y_\perp|e^{-i{p_\perp^2\over\alpha s}y_\ast}(\delta_i^j\partial_\perp^2U+2\partial_i\partial^jU)
{p_j\over p_\perp^2}|y_\perp)
\label{3.30}
\end{eqnarray}
where we put upper and lower cutoffs for the rapidity integrals, see the discussion following Eq. (\ref{TMD}). 
After Fourier transformation Eq. (\ref{3.30}) turns to
\begin{eqnarray}
&&\hspace{-1mm}
\calf_i^a(\beta_B, y_\perp)~\stackrel{\rm 2nd}{=}~
-~N_c\calf_i^a(\beta_B, y_\perp)
\!\int_{\sigma'}^\sigma\! {\dhd\alpha \over\alpha}(y_\perp|{\alpha\beta_Bs\over p_\perp^2(\alpha\beta_Bs-p_\perp^2+i\epsilon)}|y_\perp)
\nonumber\\
&&\hspace{11mm}
-~\!\int_{\sigma'}^\sigma\!{\dhd\alpha\over \alpha}
{\rm Tr}T^aU_y^\dagger(y_\perp|{1\over\alpha\beta_Bs-p_\perp^2+i\epsilon}
(\delta_i^j\partial_\perp^2U+2\partial_i\partial^jU)
{p_j\over p_\perp^2}|y_\perp)
\label{virtmast1}
\end{eqnarray}
Note that this equation can be obtained from Eq. (4.56) from Ref. \cite{gTMD1} by reversing the sign of $\beta_B$. 
In doing so one should go around the singularity at  $\alpha\beta_Bs=p_\perp^2$ according to Feynman rules since it
corresponds  to the diagram in Fig. \ref{fig:2}b with cut gluon propagator.
 
The virtual part in the complex conjugate amplitude can be similarly obtained from Eq. (4.60) from Ref. \cite{gTMD1} by 
replacement $\beta_B\rightarrow -\beta_B$. The singular denominators should look like 
${1\over\alpha\beta_Bs-p_\perp^2-i\epsilon}$ as appropriate for the  complex conjugate amplitude. 

\section{Evolution equation for gluon TMDs}

 Now we are in a position to assemble all leading-order contributions to the rapidity evolution of gluon TMDs. 
As we discussed, in  
the production part of the evolution equation for the matrix 
element (\ref{TMD}) can be obtained from Eq.  (4.28)  from Ref. \cite{gTMD1}
by replacement $\infty\leftrightarrow -\infty$. 
Adding the virtual correction to the amplitude (\ref{virtmast1}) and its complex conjugate  
we obtain the evolution equation for gluon TMD operator (\ref{operator}) in the form:
\begin{eqnarray}
&&\hspace{-1mm}
{d\over d\ln\sigma}\tilcaf_i^a(\beta_B, x_\perp) \calf_j^a(\beta_B, y_\perp)~
\label{master1alt}\\
&&\hspace{-1mm}
=~-\alpha_s{\rm Tr}\Big\{\!\int\!\dhd^2k_\perp
(x_\perp|\Big\{U^\dag{1\over\sigma\beta_Bs+p_\perp^2}
(U k_k+p_kU){\sigma \beta_Bsg_{\mu i}-2k^\perp_{\mu}k_i\over\sigma \beta_Bs+k_\perp^2}
\nonumber\\
&&\hspace{-1mm}
-~2k^\perp_\mu g_{ik}U^\dagger{1\over \sigma\beta_Bs+p_\perp^2}U -2g_{\mu k} U^\dagger{p_i\over\sigma\beta_Bs+p_\perp^2}
U+{2k^\perp_\mu\over k_\perp^2}g_{ik}\Big\}
\ticalf^k\big(\beta_B+{k_\perp^2\over\sigma s}\big)|k_\perp)
\nonumber\\
&&\hspace{5mm}
\times~(k_\perp|\calf^l\big(\beta_B+{k_\perp^2\over\sigma s}\big)
\Big\{{\sigma \beta_Bs\delta^\mu_j-2k_\perp^{\mu}k_j\over\sigma\beta_Bs+k_\perp^2}(k_lU^\dagger+U^\dagger p_l){1\over\sigma \beta_Bs+p_\perp^2}U
\nonumber\\
&&\hspace{22mm}
-2k_\perp^\mu g_{jl}U^\dagger{1\over \sigma\beta_Bs+p_\perp^2}U-~2\delta_l^\mu U^\dagger{p_j\over\sigma\beta_Bs+p_\perp^2}U
+2g_{jl}{k_\perp^\mu\over k_\perp^2}\Big\}|y_\perp)
\nonumber\\
&&\hspace{5mm}
+~2\ticalf_i(\beta_B, x_\perp)
(y_\perp|{p^m\over p_\perp^2}\calf_k(\beta_B)(i\!\stackrel{\leftarrow}{\partial}_l+U_l)(2\delta_m^k\delta_j^l-g_{jm}g^{kl})
U^\dagger{1\over \sigma\beta_Bs-p_\perp^2+i\epsilon}U
\nonumber\\
&&\hspace{77mm}
+\calf_j(\beta_B){\sigma\beta_Bs\over p^2_\perp(\sigma\beta_Bs-p_\perp^2+i\epsilon)}|y_\perp)
\nonumber\\
&&\hspace{5mm}
+~2(x_\perp|
-U^\dagger{1\over \sigma \beta_Bs-p_\perp^2-i\epsilon}U(2\delta_i^k\delta_m^l-g_{im}g^{kl} )(i\partial_k-U_k)\ticalf_l(\beta_B)
{p^m\over p_\perp^2}
\nonumber\\
&&\hspace{44mm}
+~\tilcaf_i(\beta_B)
{\sigma\beta_Bs\over p_\perp^2(\sigma\beta_Bs-p_\perp^2-i\epsilon)}|x_\perp)
\calf_j(\beta_B, y_\perp)\Big\}~+~O(\alpha_s^2)
\nonumber
\end{eqnarray}
Here the operators $\tilcaf_i(\beta)$ and 
$\calf_j(\beta)$
are defined as 
\begin{eqnarray}
&&\hspace{-1mm}
(x_\perp|\tilcaf_i(\beta)|k_\perp)
~=~{2\over s}\int\! dx_\ast ~\ticalf_i(x_\ast, x_\perp)e^{-i\beta x_\ast+i(k,x)_\perp}
\nonumber\\
&&\hspace{-1mm}
(k_\perp|\calf_i(\beta)|y_\perp)
~=~{2\over s}\int\! dy_\ast ~e^{i\beta y_\ast-i(k,y)_\perp}\calf_i(y_\ast, y_\perp)
\label{6.4}
\end{eqnarray}
Again, this equation can be reconstructed from Eq. (5.2) from Ref. \cite{gTMD1}. It should be emphasized that the reconstruction is by no means trivial: one should change 
$\infty p_1\leftrightarrow -\infty p_1$ in the production part of the amplitude 
and change $\infty p_1\leftrightarrow -\infty p_1$ and $\beta_B\leftrightarrow -\beta_B$ in the virtual part.
\footnote{
The difference between the changes in the real and virtual part of the kernel comes from the fact that in the production part we insert the full set of out-states and use double functional integral (\ref{doublefun}) afterwards. The ``total'' replacement of lightcone time 
$\infty\leftrightarrow -\infty$ would imply also the insertion of the full set of in-states. In this case the real part of the kernel will also
undergo the replacement $\beta_B\leftrightarrow-\beta_B$ leading to singularities ${1\over\alpha\beta_Bs-p^2_\perp}$ in the production part of the amplitude. In addition, there will be diagrams with both $F_{\bu i}$ and $F_\bu^{~j}$ on one side of the cut which will probably cancel these singularities. In any case, the good way to avoid these complications is to insert full set of out-states but 
use ``group law'' (\ref{grouplaw}) for $\calo$ operators to set the endpoints of gauge links to $-\infty$.}

The evolution equation (\ref{master1alt}) can be rewritten in the form where cancellation of IR and UV divergencies is evident
\begin{eqnarray}
&&\hspace{-1mm}
{d\over d\ln\sigma}\tilcaf_i^a(\beta_B,x_\perp) \calf_j^a(\beta_B,y_\perp)~
\label{master2alt}\\
&&\hspace{-1mm}
=~-\alpha_s{\rm Tr}\Big\{\!\int\!\dhd^2k_\perp
(x_\perp|\Big\{ U^\dagger{1\over\sigma\beta_Bs+p_\perp^2}
( U k_k+p_k U){\sigma \beta_Bsg_{\mu i}-2k^\perp_{\mu}k_i\over\sigma \beta_Bs+k_\perp^2}
\nonumber\\
&&\hspace{-1mm}
-~2k^\perp_\mu g_{ik} U^\dagger{1\over \sigma\beta_Bs+p_\perp^2} U
-2g_{\mu k}  U^\dagger{p_i\over\sigma\beta_Bs+p_\perp^2} U\Big\}
\ticalf^k\big(\beta_B+{k_\perp^2\over\sigma s}\big)|k_\perp)
\nonumber\\
&&\hspace{-1mm}
\times~(k_\perp|\calf^l\big(\beta_B+{k_\perp^2\over\sigma s}\big)
\Big\{{\sigma \beta_Bs\delta^\mu_j-2k_\perp^{\mu}k_j\over\sigma\beta_Bs+k_\perp^2}(k_lU^\dagger+U^\dagger p_l){1\over\sigma \beta_Bs+p_\perp^2}U
\nonumber\\
&&\hspace{-1mm}-2k_\perp^\mu g_{jl}U^\dagger{1\over \sigma\beta_Bs+p_\perp^2}U
-~2\delta_l^\mu U^\dagger{p_j\over\sigma\beta_Bs+p_\perp^2}U\Big\}|y_\perp)+~2\int\dhd^2k_\perp(x_\perp|\ticalf_i\big(\beta_B+{k_\perp^2\over\sigma s}\big)|k_\perp)
\nonumber\\
&&\hspace{-1mm}
\times(k_\perp|\calf^l\big(\beta_B+{k_\perp^2\over\sigma s}\big)
\Big\{{k_j\over k_\perp^2}{\sigma \beta_Bs+2k_\perp^2\over\sigma\beta_Bs+k_\perp^2}(k_lU^\dagger+U^\dagger p_l)
{1\over\sigma \beta_Bs+p_\perp^2}U
\nonumber\\
&&\hspace{44mm}
+~2U^\dagger{g_{jl}\over \sigma\beta_Bs+p_\perp^2}U-2{k_l\over k_\perp^2}U^\dagger{p_j\over \sigma\beta_Bs+p_\perp^2}U\Big\}
|y_\perp)
\nonumber\\
&&\hspace{-1mm}
+~2\int\dhd^2k_\perp(x_\perp|\Big\{ U^\dagger{1\over\sigma\beta_Bs+p_\perp^2}
( U k_k+p_k U){k_i\over k_\perp^2}
{\sigma \beta_Bs+2k_\perp^2\over\sigma \beta_Bs+k_\perp^2}
+2 U^\dagger{g_{ik}\over \sigma\beta_Bs+p_\perp^2} U
\nonumber\\
&&\hspace{-1mm}
-~2 U^\dagger{p_i\over\sigma\beta_Bs+p_\perp^2} U{k_k\over k_\perp^2}\Big\}
\ticalf^k\big(\beta_B+{k_\perp^2\over\sigma s}\big)|k_\perp)(k_\perp|\calf_j\big(\beta_B+{k_\perp^2\over\sigma s}\big)|y_\perp)
\nonumber\\
&&\hspace{-1mm}
+~2\ticalf_i(\beta_B, x_\perp)
(y_\perp|{p^m\over p_\perp^2}\calf_k(\beta_B)(i\!\stackrel{\leftarrow}{\partial}_l+U_l)(2\delta_m^k\delta_j^l-g_{jm}g^{kl})
U^\dagger{1\over \sigma\beta_Bs-p_\perp^2+i\epsilon}U|y_\perp)
\nonumber\\
&&\hspace{-1mm}
-~2(x_\perp|
 U^\dagger{1\over \sigma \beta_Bs-p_\perp^2-i\epsilon} 
 U(2\delta_i^k\delta_m^l-g_{im}g^{kl} )(i\partial_k- U_k)\tilde{\calf}_l(\beta_B)
{p^m\over p_\perp^2}|x_\perp)\calf_j(\beta_B, y_\perp)
\nonumber\\
&&\hspace{-1mm}
-~4\!\int\!{\dhd^2k_\perp\over k_\perp^2}\Big[\ticalf_i\big(\beta_B+{k_\perp^2\over\sigma s}, x_\perp\big)
\calf_j\big(\beta_B+{k_\perp^2\over\sigma s}, y_\perp\big)e^{i(k,x-y)_\perp}
\nonumber\\
&&\hspace{55mm}
~-{\rm V.p.}{\sigma\beta_Bs\over \sigma\beta_Bs-k_\perp^2}\ticalf_i(\beta_B, x_\perp)\calf_j(\beta_B, y_\perp)\Big]\Big\}
~+~O(\alpha_s^2)
\nonumber
\end{eqnarray}
The evolution equation (\ref{master2alt}) is one of the main results of this paper. 
It describes the rapidity evolution of the operator at any Bjorken $x_B\equiv\beta_B$ and any transverse momenta.

When we consider the evolution of gluon TMD (\ref{gTMD}) given by the matrix element (\ref{TMD}) of the operator we need 
to take into account the kinematical constraint $k_\perp^2\leq \alpha(1-\beta_B)s$ in the production part of the amplitude 
coming from the fact that matrix element $\langle\!\langle p|\ticalf_i\big(\beta_B+{k_\perp^2\over\sigma s}\big)\calf_j\big(\beta_B+{k_\perp^2\over\sigma s}\big)|p\rangle\!\rangle$ vanishes outside of this region.
(In other words,  the initial hadron's momentum is $\simeq p_2$ and the sum of 
the fraction $\beta_Bp_2$  and the fraction ${p_\perp^2\over\alpha s}p_2$ carried by the emitted gluon should be smaller than $p_2$.)
It is convenient to display this kinematical restriction explicitly so we obtain ($\eta\equiv\ln\sigma$)
\begin{eqnarray}
&&\hspace{-1mm}
{d\over d\eta}\langle\!\langle p|\tilcaf_i^a(\beta_B, x_\perp) \calf_j^a(\beta_B, y_\perp)|p\rangle\!\rangle^\eta~
\label{masterdis}\\
&&\hspace{-1mm}
=~-\alpha_s\langle\!\langle p|{\rm Tr}\Big\{\!\int\!\dhd^2k_\perp\theta\big(1-\beta_B-{k_\perp^2\over\sigma s}\big)\Big[
(x_\perp|\Big( U^\dagger{1\over\sigma\beta_Bs+p_\perp^2}
( U k_k+p_k U)
\nonumber\\
&&\hspace{-1mm}
\times~{\sigma \beta_Bsg_{\mu i}-2k^\perp_{\mu}k_i\over\sigma \beta_Bs+k_\perp^2}-~2k^\perp_\mu g_{ik} U^\dagger{1\over \sigma\beta_Bs+p_\perp^2} U
-2g_{\mu k}  U^\dagger{p_i\over\sigma\beta_Bs+p_\perp^2} U\Big)
\ticalf^k\big(\beta_B+{k_\perp^2\over\sigma s}\big)|k_\perp)
\nonumber\\
&&\hspace{-1mm}
\times~(k_\perp|\calf^l\big(\beta_B+{k_\perp^2\over\sigma s}\big)
\Big({\sigma \beta_Bs\delta^\mu_j-2k_\perp^{\mu}k_j\over\sigma\beta_Bs+k_\perp^2}(k_lU^\dagger+U^\dagger p_l){1\over\sigma \beta_Bs+p_\perp^2}U
\nonumber\\
&&\hspace{-1mm}-2k_\perp^\mu g_{jl}U^\dagger{1\over \sigma\beta_Bs+p_\perp^2}U
-~2\delta_l^\mu U^\dagger{p_j\over\sigma\beta_Bs+p_\perp^2}U\Big)|y_\perp)
\nonumber\\
&&\hspace{-1mm}
+~2(x_\perp|\ticalf_i\big(\beta_B+{k_\perp^2\over\sigma s}\big)|k_\perp)
(k_\perp|\calf^l\big(\beta_B+{k_\perp^2\over\sigma s}\big)
\Big({k_j\over k_\perp^2}{\sigma \beta_Bs+2k_\perp^2\over\sigma\beta_Bs+k_\perp^2}(k_lU^\dagger+U^\dagger p_l)
{1\over\sigma \beta_Bs+p_\perp^2}U
\nonumber\\
&&\hspace{-1mm}
+~2U^\dagger{g_{jl}\over \sigma\beta_Bs+p_\perp^2}U-2{k_l\over k_\perp^2}U^\dagger{p_j\over \sigma\beta_Bs+p_\perp^2}U
\Big)
|y_\perp)
\nonumber\\
&&\hspace{-1mm}
+~2(x_\perp|\Big( U^\dagger{1\over\sigma\beta_Bs+p_\perp^2}
( U k_k+p_k U){k_i\over k_\perp^2}
{\sigma \beta_Bs+2k_\perp^2\over\sigma \beta_Bs+k_\perp^2}
+2 U^\dagger{g_{ik}\over \sigma\beta_Bs+p_\perp^2} U
\nonumber\\
&&\hspace{-1mm}
-~2 U^\dagger{p_i\over\sigma\beta_Bs+p_\perp^2} U{k_k\over k_\perp^2}\Big)
\ticalf^k\big(\beta_B+{k_\perp^2\over\sigma s}\big)|k_\perp)(k_\perp|\calf_j\big(\beta_B+{k_\perp^2\over\sigma s}\big)|y_\perp)\Big]
\nonumber\\
&&\hspace{-1mm}
+~2\ticalf_i(\beta_B, x_\perp)
(y_\perp|{p^m\over p_\perp^2}\calf_k(\beta_B)(i\!\stackrel{\leftarrow}{\partial}_l+U_l)(2\delta_m^k\delta_j^l-g_{jm}g^{kl})
U^\dagger{1\over \sigma\beta_Bs-p_\perp^2+i\epsilon}U|y_\perp)
\nonumber\\
&&\hspace{-1mm}
-~2(x_\perp|
 U^\dagger{1\over \sigma \beta_Bs-p_\perp^2-i\epsilon} 
 U(2\delta_i^k\delta_m^l-g_{im}g^{kl} )(i\partial_k- U_k)\tilde{\calf}_l(\beta_B)
{p^m\over p_\perp^2}|x_\perp)\calf_j(\beta_B, y_\perp)
\nonumber\\
&&\hspace{-1mm}
-~4\!\int\!{\dhd^2k_\perp\over k_\perp^2}\Big[\theta\big(1-\beta_B-{k_\perp^2\over\sigma s}\big)\ticalf_i\big(\beta_B+{k_\perp^2\over\sigma s}, x_\perp\big)
\calf_j\big(\beta_B+{k_\perp^2\over\sigma s}, y_\perp\big)e^{i(k,x-y)_\perp}
\nonumber\\
&&\hspace{45mm}
~-{\rm V.p.}{\sigma\beta_Bs\over \sigma\beta_Bs-k_\perp^2}\ticalf_i(\beta_B, x_\perp)\calf_j(\beta_B, y_\perp)\Big]\Big\}
|p\rangle\!\rangle^\eta~+~O(\alpha_s^2)
\nonumber
\end{eqnarray}
 This equation describes the rapidity evolution of gluon TMD (\ref{TMD}) with rapidity cutoff (\ref{cutoff}) in the whole
range of $\beta_B=x_B$ and $k_\perp$ ($\sim|x-y|_\perp^{-1}$). In the next section we will consider some specific cases.

\section{BK, DGLAP, and Sudakov limits of TMD evolution equation}

\subsection{Small-x case: BK evolution of the Weizsacker-Williams distribution}

First, let us consider the evolution of  Weizsacker-Williams (WW) unintegrated gluon distribution 
\begin{eqnarray}
&&\hspace{-12mm}
\left.\alpha_sx_B\cald(x_B,z_\perp)\right|_{x_B\rightarrow 0}
~=~-{1\over8\pi^2(p\cdot n)}\!\int\! du \sum_X \langle p|\tilcaf^a_\xi(z_\perp+un)|X\rangle \langle X| \calf^{a\xi}(0)|p\rangle
\label{wwtmd}
\end{eqnarray}
 which can be obtained 
from Eq. (\ref{masterdis}) by setting $\beta_B=0$. Moreover, in the small-$x$ regime 
it is assumed that the energy is much higher than anything else so the characteristic transverse momenta 
$p_\perp^2\sim(x-y)_\perp^{-2}\ll s$ and in the whole range of evolution ($1\gg\sigma\gg {(x-y)_\perp^{-2}\over s}$) we have 
${p_\perp^2\over\sigma s}\ll 1$, hence the kinematical constraint $\theta\big(1-\beta_B-{k_\perp^2\over\sigma s}\big)$ in Eq. (\ref{masterdis}) can be omitted. 
Under these assumptions, all $\calf_i\big(\beta_B+{p_\perp^2\over\sigma s}\big)$ and $\calf_i(\beta_B)$ 
can be replaced by $U^\dagger i\partial_iU$  (and similarly
for $\ticalf_i$). After some algebra one obtains (cf. Eq. (6.1) from Ref. \cite{gTMD1}) 
\begin{eqnarray}
&&\hspace{-2mm}
{d\over d\ln\sigma} U^a_i(x_\perp)U^a_j(y_\perp)~
=~-4\alpha_s{\rm Tr}\Big\{\big(x_\perp\big|
 U^\dagger p_i U\big({p^k\over p_\perp^2} U^\dagger 
- U^\dagger{p^k\over p_\perp^2}\big)\big(U{p_k\over p_\perp^2} 
-{p_k\over p_\perp^2}U\big)U^\dagger p_j U|y_\perp)
\nonumber\\
&&\hspace{37mm}
-~\Big[
(x_\perp| U^\dagger{p_ip^k\over p_\perp^2} U{p_k\over p_\perp^2}|x_\perp)
-\half(x_\perp|{1\over p_\perp^2}|x_\perp) U_i(x_\perp)\Big]
U_j(y_\perp)
\nonumber\\
&&\hspace{42mm}
-~ U_i(x_\perp)\Big[
(y_\perp|{p^k\over p_\perp^2}U^\dagger{p_jp_k\over p_\perp^2}U|y_\perp)
-\half(y_\perp|{1\over p_\perp^2}|y_\perp)U_j
(y_\perp)\Big]\Big\}
\label{7.1}
\end{eqnarray}
which agrees with Ref. \cite{prd99}. This equation can be rewritten as ($\eta\equiv\ln\sigma$)
\begin{eqnarray}
&&\hspace{-1mm} 
{d\over d\eta} U^a_i(z_1) U^a_j(z_2)
\label{WWBK}\\
&&\hspace{-1mm}
=~-{g^2\over 8\pi^3}{\rm Tr}\big\{
(i\partial^{z_1}_i+ U^{z_1}_i)\big[\!\int\! d^2z_3( U^\dagger_{z_1} U_{z_3}-1)
{z_{12}^2\over z_{13}^2z_{23}^2}(U^\dagger_{z_3}U_{z_2}-1)\big]
(-i\stackrel{\leftarrow}{\partial^{z_2}_j}+U^{z_2}_j)\big\}
\nonumber 
\end{eqnarray}
where all indices are 2-dimensional and Tr stands for the trace in the adjoint representation. 
Note that the expression in the square brackets is actually the BK kernel \cite{npb96,yura}. One should also mention that
Eq. (\ref{WWBK}) coincides with Eq. (12) from Ref. \cite{domumuxi} after some algebra.

Similarly to $+\infty$ case,  the Eq. (\ref{WWBK}) holds true also at small $\beta_B$ up to 
$\beta_B\sim {(x-y)_\perp^{-2}\over s}$ since in the whole range of evolution 
$1\gg\sigma\gg{(x-y)_\perp^{-2}\over s}$ one can neglect $\sigma\beta_Bs$ in comparison to $p_\perp^2$ in Eq. (\ref{masterdis}).
This effectively reduces $\beta_B$ to 0 so one reproduces Eq. (\ref{WWBK}).

\subsection{Large transverse momenta and the light-cone limit \label{sec:6.2}}

Now let us discuss the case when $\beta_B=x_B\sim 1$ and $(x-y)_\perp^{-2}\sim s$.  At the start of the evolution
(at $\sigma\sim 1$) the cutoff in $p_\perp^2$ in the integrals Eq. (\ref{masterdis}) is $\sim(x-y)_\perp^{-2}$. However, as the evolution
in rapidity ($\sim\ln\sigma$) progresses the characteristic $p_\perp^2$ become smaller due to the kinematical constraint 
$p_\perp^2<\sigma (1-\beta_B)s$. Due to this kinematical constraint evolution in $\sigma$ is correlated with the evolution in $p_\perp^2$:
if $\sigma\gg \sigma'$ the corresponding transverse momenta of background fields ${p'_\perp}^2$ are much smaller than $p_\perp^2$ in quantum loops.
This means that during the evolution we are always in the light-cone case considered in Sect. 3 and therefore the evolution equation for
$\beta_B=x_B\sim 1$ and $(x-y)_\perp^{-2}\sim s$  takes the form
\begin{eqnarray}
&&\hspace{-1mm}
{d\over d\ln\sigma}\langle\!\langle p|\ticalf_i^a(\beta_B,x_\perp)\calf_j^a(\beta_B,y_\perp)|p\rangle\!\rangle~
\label{5.4}\\
&&\hspace{-1mm}
=~{g^2N_c\over\pi}\!\int\!\dhd^2k_\perp~\Big\{e^{i(k,x-y)_\perp}
\langle\!\langle p|\ticalf_k^a\big(\beta_B+{k_\perp^2\over\sigma s},x_\perp\big)\calf_l^a\big(\beta_B+{k_\perp^2\over\sigma s},y_\perp\big)
|p\rangle\!\rangle
\nonumber\\
&&\hspace{-1mm}
\times~\Big[{\delta_i^k\delta_j^l\over k_\perp^2}-{2\delta_i^k\delta_j^l\over\sigma\beta_Bs+k_\perp^2}
+~{k_\perp^2\delta_i^k\delta_j^l+\delta_j^kk_ik^l+\delta_i^lk_jk^k-\delta_j^lk_ik^k-\delta_i^kk_jk^l-g^{kl}k_ik_j-g_{ij}k^kk^l
\over (\sigma\beta_Bs+k_\perp^2)^2}
\nonumber\\
&&\hspace{2mm}
+~k_\perp^2{2g_{ij}k^kk^l+\delta_i^kk_jk^l+\delta_j^lk_ik^k-\delta_j^kk_ik^l-\delta_i^lk_jk^k\over (\sigma\beta_Bs+k_\perp^2)^3}
-{k_\perp^4g_{ij}k^kk^l\over  (\sigma\beta_Bs+k_\perp^2)^4}\Big]
\theta\big(1-\beta_B-{k_\perp^2\over \sigma s}\big)
\nonumber\\
&&\hspace{58mm}
-~ {\rm V.p.}{\sigma\beta_Bs\over k_\perp^2(\sigma\beta_Bs-k_\perp^2)} 
\langle\!\langle p|\ticalf_i^a(\beta_B,x_\perp)\calf_j^a(\beta_B,y_\perp)|p\rangle\!\rangle\Big\}
\nonumber
\end{eqnarray}
 which reduces to the system of 
evolution equations for  gluon TMDs $\cald(\beta_B,|z_\perp|,\ln\sigma)$ and $\calh(\beta_B,|z_\perp|,\ln\sigma)$ in the case of unpolarized hadron.
The evolution equation (\ref{5.4})
 can be rewritten as a system of evolution equations for $\cald$ and $\calh''$ functions
 ($z'\equiv {\sigma s\beta_B\over k_\perp^2+\sigma s\beta_B}$):
\begin{eqnarray}
&&\hspace{-1mm}
{d\over d\eta}\alpha_s\cald(\beta_B,z_\perp,\eta)
\label{liconefinal}\\
&&\hspace{-1mm}
=~ {\alpha_sN_c\over \pi}\!\int_{\beta_B}^1\!{dz'\over z'}\Big\{
J_0\Big(|z_\perp|\sqrt{\sigma s\beta_B{1-z'\over z'}}\Big)
\Big[\big({1\over 1-z'}\big)_+ +{1\over z'}-2+z'(1-z')\Big]\alpha_s\cald\big({\beta_B\over z'},z_\perp,\eta\big)   
\nonumber\\
&&\hspace{-1mm}
+~  
{4\over m^2}(1-z')z'
z_\perp^2J_2\Big(|z_\perp|\sqrt{\sigma s\beta_B{1-z'\over z'}}\Big)
\alpha_s\calh''({\beta_B\over z'},z_\perp,\eta)\Big\},
\nonumber\\
&&\hspace{-1mm}
{d\over d\eta}\alpha_s\calh''(\beta_B,z_\perp,\eta)
\nonumber\\
&&\hspace{-1mm}
=~ {\alpha_sN_c\over \pi}\!\int_{\beta_B}^1\!{dz'\over z'}\Big\{
J_0\Big(|z_\perp|\sqrt{\sigma s\beta_B{1-z'\over z'}}\Big)
\Big[\big({1\over 1-z'}\big)_+ -1\Big]\alpha_s\calh''\big({\beta_B\over z'},z_\perp,\eta\big)
\nonumber\\
&&\hspace{-1mm}
+~~{m^2\over 4z_\perp^2}{1-z'\over z'}J_2\Big(|z_\perp|\sqrt{\sigma s\beta_B{1-z'\over z'}}\Big)
\alpha_s\cald\big({\beta_B\over z'},z_\perp,\eta\big)\Big\}
\nonumber
\end{eqnarray}
where $\int^1_x dz f(z)g(z)_+=\int^1_x dz f(z)g(z)-\int^1_0 dz f(1)g(z)$
\footnote{
Careful analysis shows that virtual correction $\sim{\rm V.p.}{\sigma\beta_Bs\over k_\perp^2(\sigma\beta_Bs-k_\perp^2)}$ 
leads to the same $(...)_+$ prescription as the virtual correction 
$\sim{\sigma\beta_Bs\over k_\perp^2(\sigma\beta_Bs+k_\perp^2)}$  for the operator $F_{\bu i}[y_\ast,+\infty]$ so the Eq. (\ref{liconefinal}) coincides with Eq. (3.29) from Ref. \cite{gTMD1}.
}.
The above equation is our final result for the rapidity evolution of gluon TMDs (\ref{gTMD}) in the near-light-cone case.

If we take the light-cone limit $x_\perp=y_\perp$ ($\Leftrightarrow z_\perp=0$) we get the (one-loop) DGLAP equation:
\begin{equation}
\hspace{-2mm}
{d\over d\eta}\alpha_s\cald(\beta_B,0_\perp,\eta)~=~{\alpha_s\over\pi}N_c
\!\int_{\beta_B}^1\! {dz'\over z'}~
\Big[\big({1\over 1-z'}\big)_+  +{1\over z'}- 2+ z'(1-z')\Big]
\alpha_s\cald\big({\beta_B\over z'},0_\perp,\eta\big)
\label{dglap}
\end{equation}
One immediately recognizes the expression in the square brackets as gluon-gluon DGLAP kernel
(the term ${11\over 12}\delta(1-z')$ is absent since we consider the gluon light-ray operator multiplied by 
 an extra  $\alpha_s$).  

\subsection{Sudakov logarithms}

Finally, let us consider the evolution of $\cald(x_B,k_\perp,\eta=\ln\sigma)$ in the region 
where  $x_B\equiv\beta_B\sim 1$ and $k_\perp^2\sim(x-y)_\perp^{-2}\sim$ few GeV$^2$. 
In this case the integrals over $p_\perp^2$ in the production part of the kernel (\ref{masterdis}) are $\sim(x-y)_\perp^{-2}\sim k_\perp^2$   so that 
$p_\perp^2\ll\sigma\beta_Bs$ for the whole range of evolution $1>\sigma>{k_\perp^2\over s}$. 
For the same reason, the kinematical constraint $\theta\big(1-\beta_B-{p_\perp^2\over\sigma s}\big)$ 
in the last line of Eq. (\ref{masterdis}) can
be omitted and we get
\begin{eqnarray}
&&\hspace{-1mm}
{d\over d\ln\sigma}\langle\!\langle p|\tilcaf^a_i(\beta_B, x_\perp) \calf_j^a(\beta_B, y_\perp)|p\rangle\!\rangle^{\rm real}~
\label{sudreal}\\
&&\hspace{-1mm}
=~4\alpha_sN_c\!\int\!{\dhd^2 p_\perp\over p_\perp^2}e^{i(p,x-y)_\perp} 
\langle\!\langle p|\tilcaf^a_i\big(\beta_B+{p_\perp^2\over\sigma s},x_\perp\big)
\calf^a_j\big(\beta_B+{p_\perp^2\over\sigma s},y_\perp\big)|p\rangle\!\rangle
\nonumber
\end{eqnarray}
As to the virtual part
\begin{eqnarray}
&&\hspace{-1mm}
{d\over d\ln\sigma}\langle\!\langle p|\tilcaf^a_i(\beta_B, x_\perp) \calf_j^a(\beta_B, y_\perp)|p\rangle\!\rangle^{\rm virtual}~
\label{sudvirtual}\\
&&\hspace{-1mm}
=~4\alpha_sN_c\!\int\!{\dhd^2 p_\perp\over p_\perp^2}\Big[-{\rm V.p.}{\sigma \beta_Bs\over\sigma \beta_Bs-p_\perp^2}\langle\!\langle p|\tilcaf^a_i(\beta_B,x_\perp)\calf^a_j(\beta_B,y_\perp)|p\rangle\!\rangle
\Big]
\nonumber\\
&&\hspace{-1mm}
+~2\alpha_s{\rm Tr}\langle\!\langle p|(x_\perp|U^\dagger{1\over \sigma \beta_Bs-p_\perp^2-i\epsilon}U
(2\delta_i^k\delta_m^l-g_{im}g^{kl} )(i\partial_k-U_k)\ticalf_l(\beta_B){p^m\over p_\perp^2}
|x_\perp)\calf_j(\beta_B, y_\perp)
\nonumber\\
&&\hspace{-1mm}
-~\ticalf_i(\beta_B, x_\perp)
(y_\perp|{p^m\over p_\perp^2}\calf_k(\beta_B)(i\!\stackrel{\leftarrow}{\partial}_l+U_l)(2\delta_m^k\delta_j^l-g_{jm}g^{kl})
U^\dagger{1\over \sigma\beta_Bs-p_\perp^2+i\epsilon}U|y_\perp)
|p\rangle\!\rangle
\nonumber
\end{eqnarray}
the two last lines can be omitted. To prove this we follow the logic of Ref. \cite{gTMD1} and consider two cases: 
the ``light-cone case'' $l_\perp^2\ll p_\perp^2$ and the ``shock-wave'' situation when  $l_\perp^2\sim p_\perp^2$. 
It is easy to see
that in the light-cone case the two last terms in the r.h.s. of Eq. (\ref{sudvirtual}) reduce to the operators of higher collinear twist.
In the shock-wave case we need to consider two sub-cases:  if $p_\perp^2\ll \sigma\beta_Bs$ 
and $p_\perp^2\sim \sigma\beta_Bs$. In the first (sub)case the two last terms in the r.h.s. of Eq. (\ref{sudvirtual}) are
again trivially negligible in comparison to the first term in the r.h.s. of that equation. In the second (sub)case 
(when $p_\perp^2\sim \sigma\beta_Bs$)  one can expand 
the operator $\calo~\equiv~\calf_k(\beta_B)(i\!\stackrel{\leftarrow}{\partial}_l+U_l)(2\delta_m^k\delta_j^l-g_{jm}g^{kl})U^\dag$  
as $\calo(z_\perp)~=~\calo(y_\perp)+(y-z)_i\partial_i\calo(y_\perp)+...$ and get
\begin{eqnarray}
&&\hspace{-1mm} 
(y_\perp|{p^m\over p_\perp^2}\calo{1\over \sigma\beta_Bs-p_\perp^2+i\epsilon}|y_\perp)~
\nonumber\\
&&\hspace{-1mm}=~
\calo_y(y_\perp|{p^m\over p_\perp^2(\sigma\beta_Bs-p_\perp^2+i\epsilon)}|y_\perp)
+i\partial^m\calo_y(y_\perp|{1\over p_\perp^2(\sigma\beta_Bs-p_\perp^2+i\epsilon)}|y_\perp)+...
\nonumber
\end{eqnarray}
The first term in the r.h.s of this equation is obviously zero while the second is\\
 $\sim \partial_m\calo{1\over\sigma\beta_B s}\ln\sigma\beta_Bs$ 
which is $O\big({m_N^2\over\sigma\beta_Bs}\big)$ in comparison to the leading first term in the r.h.s. of Eq. (\ref{sudvirtual}) (the transverse momenta
inside the hadron target are $\sim m_N\sim 1$GeV). 

Thus, we obtain the following rapidity evolution equation in the Sudakov region:
\begin{eqnarray}
&&\hspace{-1mm}
{d\over d\ln\sigma}\langle\!\langle p|\tilcaf^a_i(\beta_B, x_\perp) \calf_j^a(\beta_B, y_\perp)|p\rangle\!\rangle~
\label{7.6}\\
&&\hspace{-1mm}
=~4\alpha_sN_c\!\int\!{\dhd^2 p_\perp\over p_\perp^2}\Big[e^{i(p,x-y)_\perp} 
\langle\!\langle p|\tilcaf^a_i\big(\beta_B+{p_\perp^2\over\sigma s},x_\perp\big)
\calf^a_j\big(\beta_B+{p_\perp^2\over\sigma s},y_\perp\big)|p\rangle\!\rangle
\nonumber\\
&&\hspace{44mm}
-~{\rm V.p.}{\sigma \beta_Bs\over\sigma \beta_Bs-p_\perp^2}\langle\!\langle p|\tilcaf^a_i(\beta_B,x_\perp)\calf^a_j(\beta_B,y_\perp)|p\rangle\!\rangle
\Big]
\nonumber
\end{eqnarray}
Similarly to Ref. \cite{gTMD1}, there is a double-log region where $1\gg\sigma\gg{(x-y)_\perp^{-2}\over s}$ 
and $\sigma\beta_Bs\gg p_\perp^2\gg (x-y)_\perp^{-2}$. In that region only the second term in the r.h.s. of 
Eq. (\ref{7.6}) survives so the evolution equation reduces to
\begin{eqnarray}
&&\hspace{-1mm} 
{d\over d\ln\sigma}\langle\!\langle p|\tilcaf^a_i(\beta_B, x_\perp)\calf^a_j(\beta_B, y_\perp)|p\rangle\!\rangle^{\eta=\ln\sigma}
\label{7.7}\\
&&\hspace{33mm} 
=~-{g^2N_c\over \pi}
\!\int\!{\dhd^2 p_\perp\over p^2_\perp}\big[1-e^{i(p,x-y)_\perp}\big]
\langle\!\langle p|\tilcaf^a_i(\beta_B, x_\perp)\calf^a_j(\beta_B, y_\perp)|p\rangle\!\rangle^\eta
\nonumber
\end{eqnarray}
which can be rewritten for the TMD (\ref{gTMD}) as
\begin{eqnarray}
&&\hspace{-1mm} 
{d\over d\ln\sigma}\cald(x_B,z_\perp,\ln\sigma)
~=~-{\alpha_sN_c\over \pi^2}\cald(x_B,z_\perp,\ln\sigma)
\!\int\!{d^2 p_\perp\over p^2_\perp}\big[1-e^{i(p,z)_\perp}\big]
\label{7.8}
\end{eqnarray}
 leading to the usual Sudakov double-log result
\begin{eqnarray}
&&\hspace{-1mm} 
\cald(x_B,k_\perp,\ln\sigma)
~\sim~\exp\big\{-{\alpha_sN_c\over 2\pi}\ln^2{\sigma s\over k_\perp^2} \big\}\cald(x_B,k_\perp,\ln{k_\perp^2\over s})
\label{sudakov}
\end{eqnarray}
It is worth noting that the coefficient in front of $\ln^2{\sigma s\over k_\perp^2}$ is determined by the
cusp anomalous dimension of two light-like Wilson lines 
going from point $y$ to $\infty p_1$ and $\infty p_2$ directions (with our cutoff $\alpha<\sigma$),
see the discussion in Ref. \cite{gTMD1}.

\section{Rapidity evolution of unintegrated gluon distribution in linear approximation}

 It is instructive to present the evolution kernel (\ref{masterdis}) in the linear (two-gluon) approximation. Since in
  the r.h.s. of Eq. (\ref{masterdis}) we already have $\ticalf_k$ and $\calf_l$ (and each of them has at least one gluon) 
 all factors $U$ and $\tilU$ in the r.h.s. of Eq. (\ref{masterdis}) can be omitted and we get ($\eta\equiv\ln\sigma$)
\begin{eqnarray}
&&\hspace{-3mm}
{d\over d\ln\sigma}\langle\!\langle p|\tilcaf_i^a(\beta_B, p_\perp) \calf_j^a(\beta_B, p'_\perp)|p\rangle\!\rangle~
\label{masterlin}\\
&&\hspace{-3mm}
=~-\alpha_sN_c\!\int\!\dhd^2k_\perp
\Big\{\theta\big(1-\beta_B-{k_\perp^2\over\sigma s}\big)
\Big[\Big({(p+k)_k\over\sigma\beta_Bs+p_\perp^2}
{\sigma \beta_Bsg_{\mu i}-2k^\perp_{\mu}k_i\over\sigma \beta_Bs+k_\perp^2}
-~2{k^\perp_\mu g_{ik}+p_ig_{\mu k} \over \sigma\beta_Bs+p_\perp^2}\Big)
\nonumber\\
&&\hspace{55mm}
\times~
\Big({\sigma \beta_Bs\delta^\mu_j-2k_\perp^{\mu}k_j\over\sigma\beta_Bs+k_\perp^2}{(p'+k)_l\over\sigma \beta_Bs+{p'}_\perp^2}
-2{k_\perp^\mu g_{jl}+\delta_l^\mu p'_j\over \sigma\beta_Bs+{p'}_\perp^2}\Big)
\nonumber\\
&&\hspace{5mm}
+~2g_{ik}\Big({k_j\over k_\perp^2}{\sigma \beta_Bs+2k_\perp^2\over\sigma\beta_Bs+k_\perp^2}
{(p'+k)_l\over\sigma \beta_Bs+{p'}_\perp^2}
+~{2g_{jl}\over \sigma\beta_Bs+{p'}_\perp^2}-{2p'_jk_l\over k_\perp^2(\sigma\beta_Bs+{p'}_\perp^2)}
\Big)
\nonumber\\
&&\hspace{5mm}
+~2g_{lj}\Big({(p+k)_k\over\sigma\beta_Bs+p_\perp^2}
{k_i\over k_\perp^2}
{\sigma \beta_Bs+2k_\perp^2\over\sigma \beta_Bs+k_\perp^2}
+{2g_{ik}\over \sigma\beta_Bs+p_\perp^2}
-~{2p_ik_k \over k_\perp^2(\sigma\beta_Bs+p_\perp^2)}\Big)\Big]
\nonumber\\
&&\hspace{52mm}
\times~
\langle\!\langle p|\ticalf^{ak}\big(\beta_B+{k_\perp^2\over\sigma s}, p_\perp-k_\perp\big)
\calf^{al}\big(\beta_B+{k_\perp^2\over\sigma s}, p'_\perp-k_\perp\big)|p\rangle\!\rangle
\nonumber\\
&&\hspace{5mm}
-~{2\over k_\perp^2}\Big[
{(2k^lp'_j-k_j{p'}^l)\delta_i^k\over \sigma\beta_Bs-(p'+k)_\perp^2+i\epsilon}
+~{(2p_ik^k-k_ip^k)\delta_j^l\over \sigma\beta_Bs-(p+k)_\perp^2-i\epsilon}\Big]
\langle\!\langle p|\ticalf_k^a(\beta_B, p_\perp)\calf_l^a(\beta_B, p'_\perp)|p\rangle\!\rangle
\nonumber\\
&&\hspace{5mm}
-~{4\over k_\perp^2}\langle\!\langle p|
\Big[\theta\big(1-\beta_B-{k_\perp^2\over\sigma s}\big)
\ticalf^a_i\big(\beta_B+{k_\perp^2\over\sigma s}, p_\perp-k_\perp\big)
\calf^a_j\big(\beta_B+{k_\perp^2\over\sigma s}, p'_\perp-k_\perp\big)
\nonumber\\
&&\hspace{55mm}
-~{\rm V.p.}{\sigma\beta_Bs\over \sigma\beta_Bs-k_\perp^2}\ticalf_i^a(\beta_B, p_\perp)\calf_j^a(\beta_B, p'_\perp)\Big]|p\rangle\!\rangle
\Big\}
\nonumber
\end{eqnarray}
where we performed Fourier transformation to the momentum space. Also, the  forward matrix element 
$\langle\!\langle p|\ticalf_i(p_\perp,\beta_B)\calf_j(p'_\perp,\beta_B)|p\rangle\!\rangle$ is proportional to $\delta^{(2)}(p_\perp-p'_\perp)$. 
Eliminating this factor and rewriting in terms of $\calr_{ij}$ (see Eq. (\ref{mael})) we obtain ($\eta\equiv\ln\sigma$)
\begin{eqnarray}
&&\hspace{-2mm}
{d\over d\eta}\calr_{ij}(\beta_B,p_\perp;\eta)
\label{linear}\\
&&\hspace{-2mm}
=~-\alpha_sN_c\!\int\!\dhd^2k_\perp\Big\{
\Big[\Big({(2p-k)_k\over\sigma\beta_Bs+p_\perp^2}
{\sigma \beta_Bsg_{\mu i}-2(p-k)^\perp_{\mu}(p-k)_i\over\sigma \beta_Bs+(p-k)_\perp^2}
-~2{(p-k)^\perp_\mu g_{ik}+p_ig_{\mu k} \over \sigma\beta_Bs+p_\perp^2}\Big)
\nonumber\\
&&\hspace{22mm}
\times~
\Big({\sigma \beta_Bs\delta^\mu_j-2(p-k)_\perp^{\mu}(p-k)_j\over\sigma\beta_Bs+(p-k)_\perp^2}
{(2p-k)_l\over\sigma \beta_Bs+p_\perp^2}
-2{(p-k)_\perp^\mu g_{jl}+\delta_l^\mu p_j\over \sigma\beta_Bs+p_\perp^2}\Big)
\nonumber\\
&&\hspace{-2mm}
+~2g_{ik}
\Big({(p-k)_j(2p-k)_l-2p_j(p-k)_l\over (p-k)_\perp^2(\sigma\beta_Bs+p_\perp^2)}
+{(p-k)_j(2p-k)_l\over (\sigma\beta_Bs+(p-k)_\perp^2)(\sigma\beta_Bs+p_\perp^2)}
+~{2g_{jl}\over \sigma\beta_Bs+p_\perp^2}\Big)
\nonumber\\
&&\hspace{-2mm}
+~2g_{lj}\Big(
{(p-k)_i(2p-k)_k-2p_i(p-k)_k\over (p-k)_\perp^2(\sigma\beta_Bs+p_\perp^2)}
+{(p-k)_i(2p-k)_k\over (\sigma\beta_Bs+(p-k)_\perp^2)(\sigma\beta_Bs+p_\perp^2)}
+{2g_{ik}\over \sigma\beta_Bs+p_\perp^2}\Big)
\Big]
\nonumber\\
&&\hspace{35mm}
\times~
\theta\big(1-\beta_B-{(p-k)_\perp^2\over\sigma s}\big)\calr^{kl}\big(\beta_B+{(p-k)_\perp^2\over\sigma s},k_\perp\big)
\nonumber\\
&&\hspace{-2mm}
-~{2\over k_\perp^2}
[\delta_i^k(k_jp^l-2k^lp_j)+\delta^l_j(k_ip^k -2p_ik^k)]{\rm V.p.}{1\over \sigma\beta_Bs-(p-k)_\perp^2}
\calr_{kl}(\beta_B,p_\perp;\eta)
\nonumber\\
&&\hspace{-2mm}
-~4\Big[{\theta\big(1-\beta_B-{(p-k)_\perp^2\over\sigma s}\big)\over (p-k)_\perp^2}
\calr_{ij}\big(\beta_B+{(p-k)_\perp^2\over \sigma s},k_\perp;\eta\big)
-{\rm V.p.}{\sigma\beta_Bs\over k_\perp^2(\sigma\beta_Bs-k_\perp^2)}\calr_{ij}(\beta_B,p_\perp;\eta)\Big]\Big\}
\nonumber
\end{eqnarray}
As we demonstrated in Ref. \cite{gTMD1} in the low-x limit $\beta_B\rightarrow 0$ the above equation reduces to the BFKL equation and the evolution of 
\begin{equation}
\beta_B\cald(\beta_B,\ln\sigma)~=~-\half\int\! \dhd^2 p_\perp \calr_i^{~i}(\beta_B,p_\perp;\ln\sigma)
\label{rii}
\end{equation}
is governed by the DGLAP equation (\ref{dglap}).

\section{Conclusions}

We have described the rapidity evolution of gluon TMD (\ref{gTMD}) with Wilson lines going to $-\infty$ in the whole range of Bjorken $x_B$ and the whole range of transverse momentum $k_\perp$. It should be emphasized that with our definition of rapidity cutoff (\ref{cutoff}) the leading-order matrix elements 
of TMD operators are UV-finite so the rapidity evolution is the only evolution and it 
describes all the dynamics of gluon TMDs (\ref{gTMD}) in the leading-log approximation.
In the next-to-leading order one should expect usual renorm-group on the top of rapidity evolution 
so the coupling constant $\alpha_s$ in our equation will become running coupling, presumably dependent
on some transverse momenta distances as in the NLO BK equation \cite{nlobk,kwa}. 

For completeness, let us present the description of various cases of linear {\it vs} nonlinear evolution 
repeating the discussion in Ref. \cite{gTMD1}.

The evolution equation  for the gluon TMD (\ref{gTMD})  with rapidity cutoff (\ref{cutoff}) is given by (\ref{masterdis}) 
and, in general, is non-linear. 
Nevertheless, for some specific cases the equation (\ref{masterdis}) linearizes. For example, let us consider the case when  $x_B\sim 1$. 
If in addition $k_\perp^2\sim s$, 
the non-linearity can be neglected for the whole range of evolution $1\gg\sigma\gg{m_N^2\over s}$ and we get the DGLAP-type system of equations (\ref{liconefinal}). 
If $k_\perp$ is small ($\sim$ few GeV) the evolution is linear
and leads to usual Sudakov factors (\ref{sudakov}). 
If we consider now the intermediate case $x_B\sim 1$ and $s\gg k_\perp^2\gg m_N^2$ the evolution
at $1\gg \sigma\gg {k_\perp^2\over s}$ will be Sudakov-type (see Eq. (\ref{7.6})) but the evolution at 
$ {k_\perp^2\over s}\gg\sigma\gg{m_N^2\over s}$ 
will be described by the full master equation (\ref{masterdis}).

For low-x region $k_\perp\sim$ few GeV and $x_B\sim{k_\perp^2\over s}$ we get the non-linear evolution described by the BK-type  equation (\ref{WWBK}).
If we now keep $k_\perp^2\sim$ few GeV$^2$ and take the intermediate $1\gg x_B\equiv\beta_B\gg {k_\perp^2\over s}$ we get a mixture of 
linear and non-linear evolutions. 
If one evolves $\sigma$ ($\leftrightarrow$ rapidity) from 1 to ${k_\perp^2\over s}$ first there will be Sudakov-type  
double-log evolution (\ref{7.8}) from $\sigma=1$ to $\sigma={k_\perp^2\over\beta_Bs}$, then the transitional region
at $\sigma\sim {k_\perp^2\over\beta_Bs}$, and after that the non-linear evolution (\ref{WWBK}) at 
${k_\perp^2\over\beta_Bs}\gg\sigma\gg{k_\perp^2\over s}$ (the interplay of the non-linear evolution and Sudakov double logarithms
in this region was studied in Ref. \cite{muxiyu} at the NLO level). The transition between the linear evolution (\ref{7.8})
and the non-linear one (\ref{WWBK}) should be described by the full equation (\ref{masterdis}). 

Another interesting case is $x_B\sim{m_N^2\over s}$ and $s\gg k_\perp^2\gg m_N^2$. In this case, if we evolve $\sigma$ from 1 to  ${m_N^2\over s}$,
first we have the BK evolution (\ref{WWBK}) up to $\sigma\sim{k_\perp^2\over s}$ and then for the evolution between $\sigma\sim{k_\perp^2\over s}$
and $\sigma\sim{m_N^2\over s}$ we need the Eq. (\ref{masterdis}) in full.

An obvious outlook project is to present the ``impact factor for the photon'' in Eq. (\ref{2.11}) for the cross section as another TMD with gauge links aligned along  
the proton's momentum. The hope is to get $k_T$-factorization in the form of product of the two TMDs 
in the whole range of Bjorken $x$ and make the
connection between $k_T$-factorization and collinear factorization.

The authors are grateful to G.A. Chirilli, J.C. Collins, Yu. Kovchegov,  A. Prokudin,  A.V. Radyushkin, T. Rogers, and F. Yuan for valuable discussions. This work was supported by contract
 DE-AC05-06OR23177 under which the Jefferson Science Associates, LLC operate the Thomas Jefferson National Accelerator Facility, and by the grant DE-FG02-97ER41028.

\section{Appendix A: Inclusive particle production as double functional integral \label{appcrsc}}
In this Section we will prove that the amplitude of inclusive particle production is given by the double functional integral (\ref{dabl}).

The cross section of the production of $\Phi$-meson in deep inelastic scattering is given by
\begin{eqnarray}
&&\hspace{-2mm}
\sigma_{\mu\nu}(x_B,s)={1\over 2\pi}\sum_X\!\int \! d^4w e^{iqw}
\langle p|j_\mu(w)|\Phi+X\rangle
\langle \Phi+X  | j_\nu(0)|p\rangle  
\label{crsc1}
\end{eqnarray}
where  $\sum_X$ denotes the sum over full set of ``out''  states.
Using standard LSZ formula we reduce Eq. (\ref{crsc1}) to
\begin{eqnarray}
&&\hspace{-2mm}
2\pi\sigma_{\mu\nu}(x_B,s)~
\nonumber\\
&&\hspace{-2mm}
=\lim_{k^2\rightarrow m^2}(k^2-m^2)^2\!\int \! d^4w d^4x d^4y e^{iqw-ikx+iky}
\sum_X\langle p|\tilT\{j_\mu(w)\Phi(x)\}|X\rangle
\langle X|T\{\Phi(y) j_\nu(0)\}|p\rangle  
\nonumber\\
&&\hspace{-2mm}
=~~\lambda^2\!\int \! d^4w d^4x d^4y e^{iqw-ikx+iky}
\sum_X\langle p|\tilT\{j_\mu(w)F^2(x)\}|X\rangle
\langle X|T\{F^2(y) j_\nu(0)\}|p\rangle  
\label{crsc2}
\end{eqnarray}
where $F^2\equiv F^m_{\alpha\beta}F^{m\alpha\beta}$ for brevity. Now, $|X\rangle$ and $|p\rangle$ may be considered as eigenstates of the full QCD Hamiltonian
\begin{equation}
\hatH|X\rangle~=~E_X|X\rangle,~~~~~~\hatH|p\rangle~=~E_p|p\rangle
\nonumber
\end{equation}
so one can rewrite $\langle X|T\{F^2(y) j_\nu(0)\}|p\rangle$ as
\begin{eqnarray}
&&\hspace{-1mm}
\langle X|T\{F^2(y) j_\nu(0)\}|p\rangle~=~e^{iE_Xt_f-iE_pt_i}\langle X|\theta(y_0)
\\
&&\hspace{-1mm}
\times~
e^{-i\hatH(t_f-y_0)}F^2(\vecy)e^{-i\hatH y_0} j_\nu(0)e^{i\hatH t_i}+~\theta(-y_0)e^{-i\hatH t_f} j_\nu(0)e^{i\hatH y_0}F^2(\vecy)e^{-i\hatH (y_0-t_i)}|p\rangle
\nonumber
\end{eqnarray}
where $t_i\rightarrow -\infty$ is the initial time and $t_f\rightarrow \infty$ is the final time.

Similarly, $\langle p|\tilT\{j_\mu(w)F^2(x)\}|X\rangle$ can be represented as
\begin{eqnarray}
&&\hspace{-1mm}
\langle p|\tilT\{j_\mu(w)F^2(x)\}|X\rangle~
\nonumber\\
&&\hspace{5mm}
=~e^{-iE_Xt_f+iE_pt_i}\langle p|\theta(w_0-x_0)e^{-i\hatH(t_i-x_0)}F^2(\vecx)e^{-i\hatH (x_0-\omega_0)} j_\nu(\vecw)e^{-i\hatH (\omega_0-t_f)}
\nonumber\\
&&\hspace{11mm}
+~\theta(x_0-w_0)e^{-i\hatH (t_i-w_0)} j_\nu(\vecw)e^{-i\hatH (w_0-x_0)}F^2(\vecx)e^{-i\hatH (x_0-t_f)}|X\rangle
\end{eqnarray}
so the cross section (\ref{crsc2}) takes the form
\begin{eqnarray}
&&\hspace{-2mm}
\sigma_{\mu\nu}(x_B,s)~=~{\lambda^2\over 2\pi}\!\int \! d^4w d^4x d^4y e^{iqw-ikx+iky}
\label{crsc4}\\
&&\hspace{-2mm}
\times~\sum_X\langle p|\theta(w_0-x_0)e^{-i\hatH(t_i-x_0)}F^2(\vecx)e^{-i\hatH (x_0-\omega_0)} j_\nu(\vecw)e^{-i\hatH (\omega_0-t_f)}
\nonumber\\
&&\hspace{22mm}
+~\theta(x_0-w_0)e^{-i\hatH (t_i-w_0)} j_\nu(\vecw)e^{-i\hatH (w_0-x_0)}F^2(\vecx)e^{-i\hatH (x_0-t_f)}|X\rangle
\nonumber\\
&&\hspace{-2mm}
\times~
\langle X|\theta(y_0)e^{-i\hatH(t_f-y_0)}F^2(\vecy)e^{-i\hatH y_0} j_\nu(0)e^{i\hatH t_i}+\theta(-y_0)e^{-i\hatH t_f} j_\nu(0)e^{i\hatH y_0}F^2(\vecy)e^{-i\hatH (y_0-t_i)}|p\rangle
\nonumber
\end{eqnarray}
At this point it is convenient to switch to the sum over all states in the ``coordinate representation'' 
\begin{equation}
\sum_X|X\rangle\langle X|~=~\int\! DA D\bsi D\psi |\vec{A}(\vecx),\psi(\vecx)\rangle \langle \vec{A}(\vecx),\psi(\vecx)|
\nonumber
\end{equation}
where $ |\vec{A}(\vecx),\psi(\vecx)\rangle$ is a state where gluon and quark fields take values  $\vec{A}$ and $\psi$ at the final time $t_f$. After this change one can rewrite
the cross section (\ref{crsc4}) in terms of the double functional integral (cf. Ref. \cite{keld})
\begin{eqnarray}
&&\hspace{-11mm}
\sigma_{\mu\nu}(x_B,s)~
=~~{\lambda^2\over 2\pi}\!\int \! d^4w d^4x d^4y e^{iqw-ikx+iky}
\int\! DA_f D\bsi_f D\psi_f 
\label{crsc5}\\
&&\hspace{-11mm}
\times~\int^{\tilA(t_f)=A_f}\! D\tilA D\tilde{\bar\psi}D\tilde{\psi}~
\Psi^\ast_p(\vec{\tilA}(t_i),\tipsi(t_i))e^{-iS_{\rm QCD}(\tilA,\tipsi)}\tilj_\mu(w)\tilF^2(x)
\nonumber\\
&&\hspace{-11mm}
\times~\!\int^{A(t_f)=A_f}DA D\bsi D\psi e^{iS_{\rm QCD}(A,\psi)}F^2(y)j_\nu(0)\Psi_p(\vec{A}(t_i),\psi(t_i))
\nonumber\\
&&\hspace{-11mm}
=~{\lambda^2\over 2\pi}\!\int \! d^4w d^4x d^4y e^{iqw-ikx+iky}
\int^{\tilA(t_f)=A(t_f)}\! D\tilA D\tilde{\bar\psi}D\tilde{\psi}DA D\bsi D\psi ~
\nonumber\\
&&\hspace{-11mm}
\times~\Psi^\ast_p(\vec{\tilA}(t_i),\tipsi(t_i))e^{-iS_{\rm QCD}(\tilA,\tipsi)}e^{iS_{\rm QCD}(A,\psi)}
\tilj_\mu(w)\tilF^2(x)F^2(y)j_\nu(0)\Psi_p(\vec{A}(t_i),\psi(t_i))
\nonumber
\end{eqnarray}
where $\Psi_p(\vec{A}(t_i),\psi(t_i))$ is the proton wave function at the initial time $t_i$.

In the same way one can demonstrate that a general matrix element  
\begin{eqnarray}
&&\hspace{-0mm} 
\langle p|\ticalo_1...\ticalo_m\calo_1...\calo_n |p'\rangle
\equiv~\sum_X\langle p| \tilT\{ \calo_1...\calo_m\}|X\rangle\langle X|T\{\calo_1...\calo_n\}|p'\rangle
\label{ourop}
\end{eqnarray}
can be represented by a double functional integral:
\begin{eqnarray}
&&\hspace{-12mm} 
\langle p|\ticalo_1...\ticalo_m\calo_1...\calo_n |p'\rangle
~=~\int\! D\tilA D\tilde{\bar\psi}D\tilde{\psi}~\Psi^\ast_p(\vec{\tilA}(t_i),\tipsi(t_i))e^{-iS_{\rm QCD}(\tilA,\tilde{\psi})}
\nonumber\\
&&\hspace{27mm} 
\times~\!\int\! DA D\bar{\psi} D\psi ~e^{iS_{\rm QCD}(A,\psi)} \ticalo_1...\ticalo_m\calo_1...\calo_n
\Psi_{p'}(\vec{A}(t_i),\psi(t_i))
\label{funtegral}
\end{eqnarray}
with the boundary condition $\tilA(\vec{x},t=\infty)=A(\vec{x},t=\infty)$ (and similarly for quark fields) reflecting the sum over all intermediate states $X$. 

\section{Appendix B: Propagators in fast background fields}

In this section we will obtain propagators for the double functional integral (\ref{3.2}) in external low-$\alpha$ fields. 
As we proved in Ref. \cite{gTMD1}, it is sufficient to consider the external field of the type $A_\bu(x_\ast,x_\perp)$ 
(and quark fields $\not\!p_1\psi(x_\ast,x_\perp)$) with all other components being zero. 
\footnote{The $z_\bu$ dependence of the external fields can be omitted since due to the rapidity ordering  $\alpha$'s of the fast fields are much less than
$\alpha$'s of the slow ones.}
Indeed, if the 
characteristic transverse momenta of fast fields ($l_\perp$) and slow fields ($k_\perp$) are comparable, 
the usual rescaling of Ref. \cite{npb96} applies so only $A_\bu(x_\ast,x_\perp)$ of the type of shock wave survives.
Conversely, if $k_\perp\gg l_\perp$ the fast fields 
do not necessarily shrink to a shock wave but we can apply the light-cone expansion
of propagators. The 
parameter of the light-cone expansion is the twist of the operator and we will expand up to operators of leading collinear twist two. Such operators
are built of two gluon operators $\sim F_{\bu i} F_{\bu j}$ or quark ones $\bsi\!\!\not\! p_1\psi$ and gauge links. 
To get coefficients in front of these operators it is sufficient to consider the external gluon field of the type $A_\bu(z_\ast,z_\perp)$ with $A_i=A_\ast=0$.

\subsection{Scalar Feynman propagator}
For simplicity we will first perform the calculation  for ``scalar propagator''  $(x|{1\over P^2+i\epsilon}|y)$. As we mentioned above, we assume that the only nonzero component of the external field is
 $A_\bu$ and it does not depend on $z_\bu$ so the operator $\alpha=i{\partial\over\partial z_\bullet}$ commutes with all background fields. 
 The propagator in the external field $A_\bu(z_\ast,z_\perp)$ has the form
\begin{eqnarray}
\hspace{-1mm}
(x|{1\over P^2+i\epsilon}|y)~&=&~\Big[-i\theta(x_\ast-y_\ast)\!\int_0^\infty\!{\dhd\alpha\over 2\alpha}
+i\theta(y_\ast-x_\ast)\!\int_{-\infty}^0\!{\dhd\alpha\over 2\alpha}\Big]~
\label{scalprop1}\\
\hspace{-1mm}
&\times&
~e^{-i\alpha(x-y)_\bu}(x_\perp|{\rm Pexp}\big\{-i\!\int_{y_\ast}^{x_\ast}\! dz_\ast
\big[{p_\perp^2\over \alpha s}-{2g\over s}A_\bu(z_\ast)\big]\big\}|y_\perp)
\nonumber
\end{eqnarray}
The Pexp in the r.h.s. of Eq. (\ref{scalprop1}) can be transformed to
\begin{eqnarray}
&&\hspace{-1mm}  
(x_\perp|e^{-i{p_\perp^2\over\alpha s}x_\ast}
{\rm Pexp}\Big\{ig\!\int^{x_\ast}_{y_\ast}\! d{2\over s}z_\ast
~e^{i{p_\perp^2\over\alpha s}z_\ast}A_\bu(z_\ast)e^{-i{p_\perp^2\over\alpha s}z_\ast}\Big\}e^{i{p_\perp^2\over\alpha s}y_\ast}
|y_\perp)~=~\int d^2z_\perp d^2z'_\perp
\label{B.12}
\\
&&\hspace{11mm}  
\times~(x_\perp|e^{-i{p_\perp^2\over\alpha s}x_\ast}|z_\perp)
(z_\perp|{\rm Pexp}\Big\{ig\!\int_{y_\ast}^{x_\ast}\! d{2\over s}z_\ast
~e^{i{p_\perp^2\over\alpha s}z_\ast}A_\bu(z_\ast)e^{-i{p_\perp^2\over\alpha s}z_\ast}\Big\}
|z'_\perp)(z'_\perp|e^{i{p_\perp^2\over\alpha s}y_\ast}
|y_\perp)
\nonumber
\end{eqnarray}
Now we expand
\begin{eqnarray}
&&\hspace{-1mm}
e^{i{p_\perp^2\over\alpha s}z_\ast}A_\bu e^{-i{p_\perp^2\over\alpha s}z_\ast}
~=~A_\bu -{z_\ast\over\alpha s}\{p^i,F_{\bu i}\}
-{z_\ast^2\over 2\alpha^2 s^2}\{p^j,\{p^i,D_jF_{\bu i}\}\}+...
\nonumber\\
&&\hspace{12mm}
=~A_\bu -{z_\ast\over\alpha s}(2p^iF_{\bu i}-iD^iF_{\bu i})
-2{z_\ast^2\over \alpha^2 s^2}(p^ip^j-ip^jD^i)D_jF_{\bu i}
+...
\label{expansion}
\end{eqnarray}
This is an expansion around the light cone $z_\perp+{2\over s}z_\ast p_1$.
We are keeping the first three terms of the expansion which is sufficient in both shock-wave case $l_\perp\sim k_\perp$
and ``light-cone'' case $l_\perp\ll k_\perp$.  In the shock-wave case it is obvious since 
the parameter of the expansion $\sim {(k,l)_\perp\over\alpha s}\sigma_\ast \ll 1$
 (recall that $\sigma_\ast\sim {\sigma s\over l_\perp^2}$).
As to the light-cone case, it is almost evident since the expansion (\ref{expansion}) gives the operators of increasing twist, 
and later we will demonstrate that three terms of the expansion are sufficient.

Using the expansion (\ref{expansion}) one easily obtains
\begin{eqnarray}
&&\hspace{-11mm}
\calo_{\alpha}(x_\ast,y_\ast)~=~{\rm Pexp}\Big\{ig\!\int_{y_\ast}^{x_\ast}\! d{2\over s}z_\ast
~e^{i{p_\perp^2\over\alpha s}z_\ast}A_\bu(z_\ast)e^{-i{p_\perp^2\over\alpha s}z_\ast}\Big\}
~=~[x_\ast,y_\ast]
\label{pa2}\\
&&\hspace{-11mm}
-~{2ig\over\alpha s^2}\!\int_{y_\ast}^{x_\ast}\!\!\!dz_\ast
~\Big(z_\ast\{p^j,[x_\ast,z_\ast]F_{\bu j}(z_\ast)[z_\ast,y_\ast]\}
+~{z_\ast^2\over 2\alpha s}\big\{p^j,\{p^k, [x_\ast,z_\ast]D_kF_{\bu j}[z_\ast,y_\ast]\}\big\}\Big)
\nonumber\\
&&\hspace{-11mm}
+~{4g^2\over\alpha s^3}\!\int_{y_\ast}^{x_\ast}\!\! \!dz_\ast\!\int_{y_\ast}^{z_\ast}\!\!\!dz'_\ast~[x_\ast,z_\ast] \Big(
-i(z-z')_\ast F_{\bu j}(z_\ast) [z_\ast,z'_\ast]F_\bu^{~j}(z'_\ast)
\nonumber\\
&&\hspace{55mm}-~4 p^jp^k{z_\ast z'_\ast\over\alpha s}F_{\bu j}(z_\ast) [z_\ast,z'_\ast]F_{\bu k}(z'_\ast)\Big)[z'_\ast,y_\ast]
+...
\nonumber
\end{eqnarray}
so the the scalar propagator in the fast external field takes the form
\begin{eqnarray}
\hspace{-1mm}
(x|{1\over P^2+i\epsilon}|y)~&=&~\Big[-i\theta(x_\ast-y_\ast)\!\int_0^\infty\!{\dhd\alpha\over 2\alpha}
+i\theta(y_\ast-x_\ast)\!\int_{-\infty}^0\!{\dhd\alpha\over 2\alpha}\Big]~e^{-i\alpha(x-y)_\bu}
\label{pa1}\\
\hspace{-1mm}
&\times&
~~(x_\perp|e^{-i{p_\perp^2\over\alpha s}x_\ast}
\calo_\alpha(x_\ast,y_\ast)e^{i{p_\perp^2\over\alpha s}y_\ast}|y_\perp)
\nonumber
\end{eqnarray}
Note that $\calo_\alpha(x_\ast,y_\ast)$ trivially satisfies the group property
\begin{equation}
\calo_\alpha(x_\ast,z_\ast)\calo_\alpha(z_\ast,y_\ast)~=~\calo_\alpha(x_\ast,y_\ast)
\label{grpr}
\end{equation}

For future use we present also two equivalent expressions with derivative operators to the right and to the left 
of the field operators:
\begin{eqnarray}
&&\hspace{-2mm}
\calo_{\alpha}(x_\ast,y_\ast)~
\nonumber\\
&&\hspace{-2mm}
=~\calo_{\alpha}(p_\perp;x_\ast,y_\ast)
=[x_\ast,y_\ast]-{2ig\over\alpha s^2}\!\int_{y_\ast}^{x_\ast}\!\!\!dz_\ast
~z_\ast\Big(2p^j[x_\ast,z_\ast]F_{\bu j}(z_\ast)-i[x_\ast,z_\ast]D^jF_{\bu j}(z_\ast)
\nonumber\\
&&\hspace{42mm}
+~2{z_\ast\over\alpha s}(p^jp^k[x_\ast,z_\ast]-ip^k[x_\ast,z_\ast]D^j)D_kF_{\bu j}\Big)[z_\ast,y_\ast]
\nonumber\\
&&\hspace{42mm}
+~{8g^2\over\alpha s^3}\!\int_{y_\ast}^{x_\ast}\!\! \!dz_\ast\!\int_{y_\ast}^{z_\ast}\!\!\!dz'_\ast~z'_\ast\Big(i
[x_\ast,z_\ast] F_{\bu j}(z_\ast) [z_\ast,z'_\ast]F_\bu^{~j}(z'_\ast)
\nonumber\\
&&\hspace{42mm}
-~2p^jp^k{z_\ast\over\alpha s}[x_\ast,z_\ast] F_{\bu j}(z_\ast) [z_\ast,z'_\ast]F_{\bu k}(z'_\ast)\Big)[z'_\ast,y_\ast]
+...
\label{repre1}\\
&&\hspace{-2mm}
=~\calo_{\alpha}(x_\ast,y_\ast;p_\perp)=[x_\ast,y_\ast]
+{2ig\over\alpha s^2}\!\int_{x_\ast}^{y_\ast}\!\!\!dz_\ast~z_\ast
~[x_\ast,z_\ast]\big\{2\tilF_{\bu j}(z_\ast)[z_\ast,y_\ast]p^j
+i\tilD^j\tilF_{\bu j}(z_\ast)[z_\ast,y_\ast]
\nonumber\\
&&\hspace{42mm}
+~2{z_\ast\over\alpha s}\big(\tilD_k\tilF_{\bu j}(z_\ast)[z_\ast,y_\ast]p^jp^k
+i\tilD^j\tilD_k\tilF_{\bu j}(z_\ast)[z_\ast,y_\ast]p^k\big)\big\}
\nonumber\\
&&\hspace{42mm}
+~{8g^2\over\alpha s^3}\!\int_{y_\ast}^{x_\ast}\!\! \!dz_\ast\!\int_{y_\ast}^{z_\ast}\!\!\!dz'_\ast~z_\ast
[x_\ast,z_\ast] \Big(
-i\tilF_{\bu j}(z_\ast) [z_\ast,z'_\ast]\tilF_\bu^{~j}(z'_\ast)[z'_\ast,y_\ast]
\nonumber\\
&&\hspace{42mm}
-~2{z'_\ast\over\alpha s}\tilF_{\bu j}(z_\ast) [z_\ast,z'_\ast]\tilF_{\bu k}(z'_\ast)[z'_\ast,y_\ast]p^jp^k\Big)
+...
\label{repre2}
\end{eqnarray}
Here we display right or left $p_\perp$ in the notation for $\calo$ to indicate   whether we use representation (\ref{repre1}) 
or  (\ref{repre2}).

To finish the proof of Eq. (\ref{pa1}) we need to demonstrate that it is correct in the light-cone case. 
We will need the general formula
\begin{eqnarray}
&&\hspace{-22mm}
\calo_\alpha^a(x_\ast,y_\ast;p_\perp)~
\label{oa}\\
&&\hspace{-22mm}
=~{\rm Pexp}\Big\{ig\!\int_{y_\ast}^{x_\ast}\! d{2\over s}z_\ast
~e^{i{p_\perp^2\over\alpha s}(z-a)_\ast}A_\bu(z_\ast)e^{-i{p_\perp^2\over\alpha s}(z-a)_\ast}\Big\}
\nonumber\\
&&\hspace{-1mm}
=~[x_\ast,y_\ast]-~{2ig\over\alpha s^2}\!\int_{y_\ast}^{x_\ast}\!\!\!dz_\ast
~\big((z-a)_\ast\{p^j,[x_\ast,z_\ast]F_{\bu j}(z_\ast)[z_\ast,y_\ast]\}
\nonumber\\
&&\hspace{27mm}
+~{(z-a)_\ast^2\over 2\alpha s}\big\{p^j,\{p^k, [x_\ast,z_\ast]D_kF_{\bu j}[z_\ast,y_\ast]\}\big\}\big)
\nonumber\\
&&\hspace{-22mm}
+~{4g^2\over\alpha s^3}\!\int_{y_\ast}^{x_\ast}\!\! \!dz_\ast\!\int_{y_\ast}^{z_\ast}\!\!\!dz'_\ast~[x_\ast,z_\ast] \big(
-i(z-z')_\ast F_{\bu j}(z_\ast) [z_\ast,z'_\ast]F_\bu^{~j}(z'_\ast)
\nonumber\\
&&\hspace{-1mm}
-~4 p^jp^k{(z-a)_\ast (z'-a)_\ast\over\alpha s}F_{\bu j}(z_\ast) [z_\ast,z'_\ast]F_{\bu k}(z'_\ast)\big)[z'_\ast,y_\ast]
+...
\nonumber
\end{eqnarray}

In the light-cone case one expands the external field either around the light cone $y_\perp+{2\over s}z_\ast p_1$ or
$x_\perp+{2\over s}z_\ast p_1$. Let us consider the first case (the second is equivalent).  
The Pexp in the r.h.s. of Eq. (\ref{scalprop1}) can be transformed to
\begin{eqnarray}
&&\hspace{-11mm}
(x_\perp|{\rm Pexp}\big\{-i\!\int_{y_\ast}^{x_\ast}\! dz_\ast\big[{p_\perp^2\over \alpha s}-{2g\over s}A_\bu(z_\ast)\big]\big\}|y_\perp)
\nonumber\\
&&\hspace{-11mm}
=~(x_\perp|e^{-i{p_\perp^2\over\alpha s}(x_\ast-y_\ast)}
{\rm Pexp}\Big\{{2ig\over s}\!\int_{y_\ast}^{x_\ast}\! dz_\ast
~e^{i{p_\perp^2\over\alpha s}(z_\ast-y_\ast)}A_\bu(z_\ast)e^{-i{p_\perp^2\over\alpha s}(z_\ast-y_\ast)}
\Big\}|y_\perp)
\label{A.2}
\end{eqnarray}
Now we rewrite Eq. (\ref{oa}) in the form (\ref{repre1})
\begin{eqnarray}
&&\hspace{-1mm}
\calo_\alpha^{y_\ast}(p_\perp; x_\ast,y_\ast)~
~=~{\rm Pexp}\Big\{ig\!\int_{y_\ast}^{x_\ast}\! d{2\over s}z_\ast
~e^{i{p_\perp^2\over\alpha s}(z-y)_\ast}A_\bu(z_\ast)e^{-i{p_\perp^2\over\alpha s}(z-y)_\ast}\Big\}
\nonumber\\
&&\hspace{18mm}
=~[x_\ast,y_\ast]-{2ig\over\alpha s^2}\!\int_{y_\ast}^{x_\ast}\!\!\!dz_\ast
~(z-y)_\ast\Big(2p^j[x_\ast,z_\ast]F_{\bu j}(z_\ast)-i[x_\ast,z_\ast]D^jF_{\bu j}(z_\ast)
\nonumber\\
&&\hspace{27mm}
+~2{(z-y)_\ast\over\alpha s}(p^jp^k[x_\ast,z_\ast]-ip^k[x_\ast,z_\ast]D^j)D_kF_{\bu j}\Big)[z_\ast,y_\ast]
\nonumber\\
&&\hspace{27mm}
+~{8g^2\over\alpha s^3}\!\int_{y_\ast}^{x_\ast}\!\! \!dz_\ast\!\int_{y_\ast}^{z_\ast}\!\!\!dz'_\ast~(z'-y)_\ast\Big(i
[x_\ast,z_\ast] F_{\bu j}(z_\ast) [z_\ast,z'_\ast]F_\bu^{~j}(z'_\ast)
\nonumber\\
&&\hspace{27mm}
-~2p^jp^k{(z-y)_\ast\over\alpha s}[x_\ast,z_\ast] F_{\bu j}(z_\ast) [z_\ast,z'_\ast]F_{\bu k}(z'_\ast)\Big)[z'_\ast,y_\ast]
+...
\label{oyast}
\end{eqnarray}
This is effectively expansion around the light ray $y_\perp+{2\over s}y_\ast p_1$ with
the parameter of the expansion $\sim {|l_\perp|\over |p_\perp|}\ll 1$. As we mentioned, we 
 expand up to the operators of twist two. Using Eq. (\ref{oyast}) we obtain
the propagator (\ref{scalprop1}) 
in the form
\begin{eqnarray}
&&\hspace{-1mm}
(x|{1\over P^2+i\epsilon}|y)~=~\Big[-i\theta(x_\ast-y_\ast)\!\int_0^\infty\!{\dhd\alpha\over 2\alpha}
+i\theta(y_\ast-x_\ast)\!\int_{-\infty}^0\!{\dhd\alpha\over 2\alpha}\Big]
\label{formula8}\\
&&\hspace{32mm}
\times~e^{-i\alpha(x-y)_\bu}(x_\perp|e^{-i{p_\perp^2\over\alpha s}(x-y)_\ast}\calo_\alpha^{y_\ast}(x_\ast,y_\ast;p_\perp)|y_\perp)
\nonumber
\end{eqnarray}
which coincides with the light-cone expansion of scalar propagator (A.6) from Ref. \cite{gTMD1}.
Thus,   the Eq. (\ref{formula8}) agrees with Eq. (\ref{pa1}).

Similarly, one can demonstrate that the propagator in the complex conjugate amplitude has the form
\begin{eqnarray}
\hspace{-1mm}
(x|{1\over P^2-i\epsilon}|y)~&=&~\Big[i\theta(y_\ast-x_\ast)\!\int_0^\infty\!{\dhd\alpha\over 2\alpha}
-i\theta(x_\ast-y_\ast)\!\int_{-\infty}^0\!{\dhd\alpha\over 2\alpha}\Big]~e^{-i\alpha(x-y)_\bu}
\label{pa2cc}\\
\hspace{-1mm}
&\times&
~~(x_\perp|e^{-i{p_\perp^2\over\alpha s}x_\ast}
\calo_\alpha(x_\ast,y_\ast)e^{i{p_\perp^2\over\alpha s}y_\ast}|y_\perp)
\nonumber
\end{eqnarray}
After transformation $e^{-i{p_\perp^2\over\alpha s}x_\ast}
\calo_\alpha(x_\ast,y_\ast)e^{i{p_\perp^2\over\alpha s}x_\ast}~=~\calo_\alpha^{x_\ast}(x_\ast,y_\ast;p_\perp)$
and rewriting according to Eq. (\ref{repre2})  this equation coincides with Eq. (A.12) from Ref. \cite{gTMD1}.

\subsection{Scalar propagator of Wightman type}
The scalar propagator from point $x$ to the left of the cut to point $y$ to the right of the cut 
reads
\begin{eqnarray}
&&\hspace{-11mm}
(x|{1\over P^2-i\epsilon}p^22\pi\delta(p^2)\theta(p_0)p^2{1\over P^2+i\epsilon}|y)
\label{scalpropmp}
\end{eqnarray}
It is convenient to represent this equation as an integral of product of two amplitudes of particle emission 
found in Ref. \cite{gTMD1}:
\begin{eqnarray}
&&\hspace{-1mm}
\lim_{k^2\rightarrow 0}k^2(k|{1\over P^2+i\epsilon}|y_\perp,y_\ast)
~=~(k_\perp|\calo_\alpha(k_\perp;\infty,y_\ast)e^{i{p_\perp^2\over\alpha s}y_\ast}|y_\perp)
\nonumber\\
&&\hspace{-1mm}
\lim_{k^2\rightarrow 0}k^2(x_\perp,x_\ast|{1\over P^2-i\epsilon}|k)~
=~(x_\perp|e^{-i{p_\perp^2\over\alpha s}x_\ast}\calo_\alpha(x_\ast,\infty;k_\perp)|k_\perp)
\label{spw2}
\end{eqnarray}
In the shock-wave case $l_\perp\sim k_\perp$ these formulas coincide with Eqs. (B.18) and (B.20)
from  Ref. \cite{gTMD1}; in the light-cone case one needs to rewrite them as
\begin{eqnarray}
&&\hspace{-1mm}
\lim_{k^2\rightarrow 0}k^2(k|{1\over P^2+i\epsilon}|y_\perp,y_\ast)
~=~e^{i{k_\perp^2\over\alpha s}y_\ast}(k_\perp|\calo_\alpha^{y_\ast}(k_\perp;\infty,y_\ast)|y_\perp)
\nonumber\\
&&\hspace{-1mm}
\lim_{k^2\rightarrow 0}k^2(x_\perp,x_\ast|{1\over P^2-i\epsilon}|k)~
=~e^{-i{k_\perp^2\over\alpha s}x_\ast}(x_\perp|\calo_\alpha^{x_\ast}(x_\ast,\infty;k_\perp)|k_\perp)
\end{eqnarray}
after which they coincide with Eqs. (A.14) and (A.16) from  Ref. \cite{gTMD1}.

Using Eq. (\ref{spw2}) one easily obtains
\begin{eqnarray}
&&\hspace{-11mm}    
(x|{1\over P^2-i\epsilon}p^22\pi\delta(p^2)\theta(p_0)p^2{1\over P^2+i\epsilon}|y)~
\nonumber\\
&&\hspace{-1mm}=~
\int_0^\infty\!{\dhd\alpha\over 2\alpha}~e^{-i\alpha(x-y)_\bu}
(x_\perp|e^{-i{p_\perp^2\over\alpha s}x_\ast}
\ticalo_\alpha(x_\ast,\infty)\calo_\alpha(\infty,y_\ast)
e^{i{p_\perp^2\over\alpha s}y_\ast}|y_\perp)
\label{scalpromp}
\end{eqnarray}
where $\ticalo$ is built of the $\tilA$ fields in the left functional integral in 
Eq. (\ref{funtegral}).

\subsection{Gluon propagator  in the light-like gauge \label{app:gluprop}}

The general expression for Feynman gluon propagator in the light-like gauge $p_2^\mu A_\mu=0$ in the background field (\ref{klfild}) 
has the form  
\begin{eqnarray}
&&\hspace{-11mm}
i\langle T\{A^a_\mu(x)A^b_\nu(y)\}\rangle
~=~(x|\big(g^\perp_{\mu i}-{p_{2\mu}\over p_\ast}p_i\big){1\over P^2+i\epsilon}
\big(\delta_\nu^i-p^i{p_{2\nu}\over p_\ast}\big)-{p_{2\mu}p_{2\nu}\over p_\ast^2}|y)^{ab}
\label{glupropaxi}
\end{eqnarray}
Using the expression (\ref{pa1}) for ${1\over P^2+i\epsilon}$ we get
\begin{eqnarray}
&&\hspace{-1mm}
\langle T\{A^a_\mu(x)A^b_\nu(y)\} \rangle
~=~\Big[-\theta(x_\ast-y_\ast)\!\int_0^\infty\!{\dhd\alpha\over 2\alpha}
+\theta(y_\ast-x_\ast)\!\int_{-\infty}^0\!{\dhd\alpha\over 2\alpha}\Big]e^{-i\alpha(x-y)_\bu}
\label{glupropax}\\
&&\hspace{-1mm}
\times~
(x_\perp|e^{-i{p_\perp^2\over\alpha s}x_\ast}\big(g^\perp_{\mu i}-{2p_{2\mu}\over \alpha s}p_i\big)
\calo_\alpha(x_\ast,y_\ast;p_\perp)\big(\delta_\nu^i-p^i{2p_{2\nu}\over \alpha s}
\big)e^{i{p_\perp^2\over\alpha s}y_\ast}|y_\perp)^{ab}+i(x|\frac{p_{2\mu} p_{2\nu}}{p^2_\ast}|y)^{ab}
\nonumber
\end{eqnarray}
For the complex conjugate amplitude one obtains in a similar way
\begin{eqnarray}
&&\hspace{-11mm}
-i\langle \tilde{T}\{A^a_\mu(x)A^b_\nu(y)\}\rangle
~=~(x|\big(g^\perp_{\mu i}-{p_{2\mu}\over p_\ast}p_i\big){1\over P^2-i\epsilon}
\big(\delta_\nu^i-p^i{p_{2\nu}\over p_\ast}\big)-{p_{2\mu}p_{2\nu}\over p_\ast^2}|y)^{ab}
\label{glupropaxicc}
\end{eqnarray}
and 
\begin{eqnarray}
&&\hspace{-1mm}
\langle \tilT\{A^a_\mu(x)A^b_\nu(y)\} \rangle
~=~\Big[-\theta(y_\ast-x_\ast)\!\int_0^\infty\!{\dhd\alpha\over 2\alpha}
+\theta(x_\ast-y_\ast)\!\int_{-\infty}^0\!{\dhd\alpha\over 2\alpha}\Big]e^{-i\alpha(x-y)_\bu}
\label{glupropax}\\
&&\hspace{-1mm}
\times~
(x_\perp|e^{-i{p_\perp^2\over\alpha s}x_\ast}\big(g^\perp_{\mu i}-{2p_{2\mu}\over \alpha s}p_i\big)
\calo_\alpha(x_\ast,y_\ast;p_\perp)\big(\delta_\nu^i-p^i{2p_{2\nu}\over \alpha s}
\big)e^{i{p_\perp^2\over\alpha s}y_\ast}|y_\perp)^{ab}-i(x|\frac{p_{2\mu}p_{2\nu}}{p^2_\ast}|y)^{ab}
\nonumber
\end{eqnarray}
where we used Eq. (\ref{pa2cc}) for ${1\over P^2-i\epsilon}$.

The ``cut'' propagator in the background field (\ref{klfild}) is given by Eq. (\ref{glupropaxmp})
\begin{eqnarray}
&&\hspace{-1mm}
\langle \tilA^a_\mu(x)A^b_\nu(y)\rangle
\nonumber\\
&&\hspace{-1mm}
~=~-(x|\big(g^\perp_{\mu i}-{p_{2\mu}\over p_\ast}p_i\big){1\over P^2-i\epsilon}p^2
2\pi\delta(p^2)\theta(p_0)p^2{1\over P^2+i\epsilon}
\big(\delta_\nu^i-p^i{p_{2\nu}\over p_\ast}\big)|y)^{ab}
\label{glupropaximp}
\end{eqnarray}
Using Eq. (\ref{scalpromp}) for  scalar propagator we obtain
\begin{eqnarray}
&&\hspace{-1mm}
\langle \tilA^a_\mu(x)A^b_\nu(y)\rangle=~
-\!\int_0^\infty\!{\dhd\alpha\over 2\alpha}~e^{-i\alpha(x-y)_\bu}
\label{glupromp}\\
&&\hspace{-1mm}
\times(x_\perp|\big(g^\perp_{\mu i}-{2p_{2\mu}\over \alpha s}p_i\big)e^{-i{p_\perp^2\over\alpha s}x_\ast}
\ticalo(x_\ast,\infty)\calo(\infty,y_\ast)
e^{i{p_\perp^2\over\alpha s}y_\ast}\big(\delta_\nu^i-p^i{2p_{2\nu}\over \alpha s}\big)|y_\perp)^{ab}
\nonumber
\end{eqnarray}
where, as usual, $\ticalo$ is built of the $\tilA$ fields in the left functional integral in 
Eq. (\ref{funtegral}).

\section{Appendix C: Feynman diagrams for the gluon propagator in the light-like gauge}

The formulas (\ref{glupropaxi}) and (\ref{glupropaxicc}) can be easily obtained from general formula for the
propagator in the light-like gauge in Ref. \cite{npb96}. However, the expression (\ref{glupropaximp}) for Wightman gluon propagator needs derivation
and the easiest way is to analyze Feynman diagrams  in the background field (\ref{klfild}) (cf.  Ref. \cite{balbel}).
\begin{figure}[htb]
\begin{center}
\includegraphics[width=126mm]{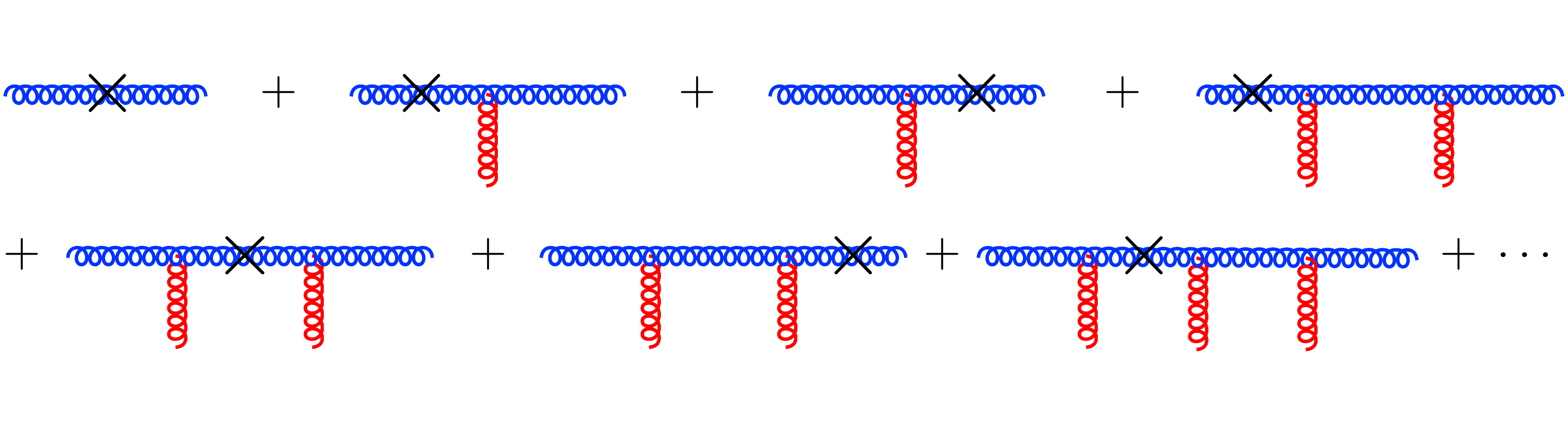}
\end{center}
\caption{Cut gluon propagator in external field $A_\bullet(x_\ast, x_\perp)$.\label{fig:3}}
\end{figure}

Let us consider a typical diagram shown in Fig. \ref{fig:3}.  The perturbative gluon propagators in the
light-like  $p_2^\mu A_\mu=0$  gauge has the form
\begin{eqnarray}
&&\hspace{-1mm}
\langle {\rm T}\{A_\mu(x)A_\nu(y)\}\rangle~=~\int\!{\dhd^4 k\over i}~{d_{\mu\nu}(k)\over k^2+i\epsilon} ~e^{-ik(x-y)},
\nonumber\\
&&\hspace{-1mm}
\langle {\rm \tilde{T}}\{A_\mu(x)A_\nu(y)\}\rangle~=~i\int\!\dhd^4 k~{d_{\mu\nu}(k)\over k^2-i\epsilon} ~e^{-ik(x-y)}  ,
\nonumber\\
&&\hspace{-1mm}
\langle \tilde{A}_\mu(x)A_\nu(y)\rangle~=~-\!\int\!\dhd^4 k~2\pi\delta(k^2)\theta(\alpha)d_{\mu\nu}(k)~e^{-ik(x-y)}
\nonumber
\end{eqnarray}
where 
\begin{equation}
d_{\mu\nu}(k)~=~g_{\mu\nu}^\perp-{2\over\alpha s}(p_{2\mu}k^\perp_\nu+p_{2\nu}k^\perp_\mu)-{4\beta\over\alpha s}p_{2\mu}p_{2\nu}
\end{equation}
First, we prove that only one term in the three-gluon vertex survives. 
Indeed, consider a typical 3-gluon vertex 
\begin{eqnarray}
&&\hspace{-1mm}
(2k+q)\cdot A(q) g_{\mu\nu}-(k+2q)_\mu A_\nu(q)+(q-k)_\nu A_\mu(q)
\nonumber\\
&&\hspace{-1mm}
=~(2k+q)\cdot A(q) g_{\mu\nu}+\frac{2}{s}[(q-k)_\nu p_{2\mu}-(k+2q)_\mu p_{2\nu}]A_\bu(q)
\nonumber
\end{eqnarray}
It is easy to see that the two last terms do not contribute since the vertex is multiplied by $d_{\alpha\mu}(k)$ and 
$d_{\nu\beta}(k+q)$ so we are left with the first term which is a vertex of emission of the gluon by scalar propagator
multiplied by $g_{\mu\nu}$.

Second, let us consider the product of numerators of gluon propagators in Fig. \ref{fig:3}
\begin{equation}
d_{\alpha\mu_1}(k)d_{\mu_1\mu_2}(k+q_1)d_{\mu_2\mu_3}(k+q_1+q_2)\dots d_{\mu_n\beta}(k+q_1+\dots+q_n)
\end{equation}
It is clear that for all $d_{\mu\nu}$'s, except the first and the last ones, we can replace $d_{\mu\nu}(k)$ by $g_{\mu\nu}^\perp$ since terms
$\sim p_{2\mu}$ vanish. For the same reason, only two terms in the first and in the last $d_{\mu\nu}$'s survive:
\begin{eqnarray}
&&\hspace{-1mm}
d_{\alpha\mu_1}(k)~\rightarrow~g^\perp_{\alpha\mu_1}-{2\over\alpha s}p_{2\alpha}k^\perp_{\mu_1},   
\nonumber\\
&&\hspace{-1mm}
d_{\mu_n\beta}(k+q_1+\dots+q_n)\rightarrow~g^\perp_{\mu_n\beta}-{2\over\alpha s}p_{2\beta}(k^\perp+q^\perp_1+\dots+q^\perp_n)_{\mu_n}
\label{twofacts}
\end{eqnarray}
Thus, the gluon propagator in the background field (\ref{klfild}) in the light-like $p_2^\mu A_\mu=0$ gauge differs from the scalar propagator
in the same background field (\ref{scalpropmp}) only by two factors (\ref{twofacts})
\begin{eqnarray}
&&\hspace{-1mm}
\langle \tilA^a_\mu(x)A^b_\nu(y)\rangle
\nonumber\\
&&\hspace{-1mm}
~=~-(x|\big(g^\perp_{\mu i}-{p_{2\mu}\over p_\ast}p_i\big){1\over P^2-i\epsilon}p^2
2\pi\delta(p^2)\theta(p_0)p^2{1\over P^2+i\epsilon}
\big(\delta_\nu^i-p^i{p_{2\nu}\over p_\ast}\big)|y)^{ab}
\label{glupropaxmp}
\end{eqnarray}

\section{Appendix D: Light-like {\rm vs} background-Feynman gauge \label{app:compare}}
In this Section we prove that our expression (\ref{masterplus}), obtained in the light-like gauge
agrees with the results of Ref. \cite{gTMD1} obtained in the background-Feynman gauge. 
First, we rewrite Eq. (\ref{masterplus}) as 
 a product of two Lipatov vertices of gluon emission 
\begin{eqnarray}
&&\hspace{-1mm}
\langle p| F^m_{\bu i}(x_\ast,x_\perp)[x_\ast,\infty]^{ma}[\infty,y_\ast]^{an}F^n_{\bu j}(y_\ast,y_\perp)|p\rangle^\eta
\nonumber\\
&&\hspace{-1mm}
=~-\!\int_{\sigma'}^\sigma\!{\dhd\alpha\over 2\alpha}\!\int\!\dhd^2 k_\perp 
\langle p| L^{ba}_{ik}(x_\perp,k_\perp; x_\ast)L_j^{k,ab}(y_\perp,k_\perp; y_\ast)|p\rangle
\label{masterplu}
\end{eqnarray}
where 
\begin{eqnarray}
&&\hspace{-1mm}
L_j^{k,ab}(y_\perp,k_\perp; y_\ast)\equiv~\lim_{k^2\rightarrow 0}k^2\langle A^{a k}(k)[\infty,y_\ast]_y^{bm} F_{\bu j}^m(y_\ast,y_\perp)\rangle
\nonumber\\
&&\hspace{-1mm}
=~{1\over 2\alpha}
\Big[(k_\perp|\calo_\alpha(\infty,y_\ast,p_\perp)e^{i{p_\perp^2\over\alpha s}y_\ast}(p_\perp^2\delta_j^k+2p_jp^k)|y_\perp)
[y_\ast,\infty]_y
\nonumber\\
&&\hspace{-1mm}
-~{4\over s}(k_\perp|\calo_\alpha(\infty,y'_\ast,p_\perp)\!\int_{y_\ast}^\infty \! dy'_\ast~e^{i{p_\perp^2\over\alpha s}y'_\ast}p^k|y_\perp)
[y'_\ast,y_\ast]_yF_{\bu j}(y_\ast,y_\perp)[y_\ast,\infty]_y\Big]^{ab}
\label{newlvertex}
\end{eqnarray}
and similarly for $L^{ba}_{ik}(x_\perp,k_\perp; x_\ast)$.

We will prove that the Lipatov vertex (\ref{newlvertex}) coincides with 
\begin{eqnarray}
&&\hspace{-1mm}
L_j^{k, ab}(y_\perp, k_\perp; y_\ast)
\nonumber\\
&&\hspace{-1mm}
=~{\theta(y_\ast)\delta^{ab}\over 2\alpha}e^{i{k_\perp^2\over\alpha s}y_\ast-i(k,y)_\perp}(k_\perp^2\delta_j^k+2k_jk^k)+{\theta(-y_\ast)\over 2\alpha}
(k_\perp|Ue^{i{p_\perp^2\over\alpha s}y_\ast}(p_\perp^2\delta_j^k+2p_jp^k)U^\dagger|y_\perp)^{ab}
\nonumber\\
&&\hspace{-1mm}
+~{1\over 2\alpha}e^{i{k_\perp^2\over\alpha s}y_\ast-i(k,y)_\perp}
\Big\{
-~{4ig\over\alpha s^2}(k_\perp^2\delta_j^k+2k_jk^k)k^l\!\int\!dz_\ast
~\big((z-y)_\ast\theta(z_\ast-y_\ast)
\nonumber\\
&&\hspace{75mm}
+y_\ast\theta(-y_\ast)\big)[\infty,z_\ast]F_{\bu l}(z_\ast)[z_\ast,\infty]
\nonumber\\
&&\hspace{-1mm}
+~(\delta_j^k k^l-g^{kl}k_j-\delta_j^lk^k)
{4\over s}\!\int\!dz_\ast~[\theta(z_\ast-y_\ast)-\theta(-y_\ast)]
~[\infty,z_\ast]F_{\bu l}(z_\ast)[z_\ast,\infty]
\Big\}^{ab}
\nonumber\\
&&\hspace{-1mm}
-~2i{k^k\over k_\perp^2}e^{i{k_\perp^2\over\alpha s}y_\ast-i(k,y)_\perp}
[\infty,y_\ast]_yF_{\bu j}(y_\ast,y_\perp)[y_\ast,\infty]_y^{ab}
\label{lvertexcoord}
\end{eqnarray}
with our accuracy.  

\subsection{Light-cone case}
Let us start with the ``light-cone case'' when the characteristic transverse momenta
of background field $l_\perp$ are much smaller than the momenta of the ``quantum'' fields $p_\perp$. 
As we discussed above, we need to find the Lipatov vertex with twist-one accuracy which 
means taking into account only first term in the expansion in powers of $F_{\bu i}$. First, let us note that 
in such approximation the last terms in Eqs. (\ref{newlvertex}) and (\ref{lvertexcoord}) coincide so we need to prove 
that
\begin{eqnarray}
&&\hspace{-1mm}
{1\over 2\alpha}(k_\perp|\calo_\alpha(\infty,y_\ast,p_\perp)e^{i{p_\perp^2\over\alpha s}y_\ast}(p_\perp^2\delta_j^k+2p_jp^k)|y_\perp)
[y_\ast,\infty]_y
\nonumber\\
&&\hspace{-1mm}
=~
{\theta(y_\ast)\over 2\alpha}(k_\perp^2\delta_j^k+2k_jk^k)e^{i{k_\perp^2\over\alpha s}y_\ast-i(k,y)_\perp}
+{\theta(-y_\ast)\over 2\alpha}
(k_\perp|Ue^{i{p_\perp^2\over\alpha s}y_\ast}(p_\perp^2\delta_j^k+2p_jp^k)U^\dagger|y_\perp)
\nonumber\\
&&\hspace{-1mm}
+~{1\over 2\alpha}e^{i{k_\perp^2\over\alpha s}y_\ast-i(k,y)_\perp}
\Big\{
-~{4ig\over\alpha s^2}(k_\perp^2\delta_j^k+2k_jk^k)k^l
\nonumber\\
&&\hspace{-1mm}
\times~\!\int\!dz_\ast
~\big((z-y)_\ast\theta(z_\ast-y_\ast)+y_\ast\theta(-y_\ast)\big)[\infty,z_\ast]F_{\bu l}(z_\ast)[z_\ast,\infty]
\nonumber\\
&&\hspace{-1mm}
+~(\delta_j^k k^l-g^{kl}k_j-\delta_j^lk^k)
{4\over s}\!\int\!dz_\ast~[\theta(z_\ast-y_\ast)-\theta(-y_\ast)]
~[\infty,z_\ast]F_{\bu l}(z_\ast)[z_\ast,\infty]
\Big\}
\label{niid2pruv}
\end{eqnarray}
Using formulas
\begin{eqnarray}
&&\hspace{-11mm}
[\infty,y_\ast]p_\perp^2[y_\ast,\infty]~=~p_\perp^2+2p^i\!\int_{y_\ast}^\infty\! d{2\over s}y'_\ast 
[\infty,y'_\ast]F_{\bu i}(y'_\ast)[y'_\ast,\infty]~+~O(DF,F^2)
\label{13.6}\\
&&\hspace{-11mm}
[\infty,y_\ast]2p_jp_k[y_\ast,\infty]~=~2p_jp_k
-2\!\int_{y_\ast}^\infty\! d{2\over s}y'_\ast 
[\infty,y'_\ast](p_jF_{\bu k}(y'_\ast)+j\leftrightarrow k)[y'_\ast,\infty]~+~O(DF,F^2)
\nonumber
\end{eqnarray}
we obtain
\begin{eqnarray}
&&\hspace{-11mm}
(k_\perp|\calo_\alpha(\infty,y_\ast,p_\perp)e^{i{p_\perp^2\over\alpha s}y_\ast}(p_\perp^2\delta_j^k+2p_jp^k)|y_\perp)
[y_\ast,\infty]_y
\nonumber\\
&&\hspace{-11mm}
=~e^{i{k_\perp^2\over\alpha s}y_\ast-i(k,y)_\perp}
\Big\{(k_\perp^2\delta_j^k+2k_jk^k)\Big(1
-~{4ig\over\alpha s^2}k^i\!\int_{y_\ast}^{\infty}\!\!\!dz_\ast
~(z-y)_\ast[\infty,z_\ast]F_{\bu i}(z_\ast)[z_\ast,\infty]\Big)
\nonumber\\
&&\hspace{-11mm}
+~(\delta_j^k k^i-g^{ik}k_j-\delta_j^ik^k)
{4\over s}\!\int_{y_\ast}^{\infty}\!\!\!dz_\ast
~[\infty,z_\ast]F_{\bu i}(z_\ast)[z_\ast,\infty]
\Big\}
\label{13.7}
\end{eqnarray}
Also, using Eqs. (\ref{13.6}) and the commutator 
$$
e^{-i{p_\perp^2\over\alpha s}y_\ast}Ue^{i{p_\perp^2\over\alpha s}y_\ast}-U\simeq -{2y_\ast\over\alpha s}k^l\partial_lU
$$ 
one finds
\begin{eqnarray}
&&\hspace{-12mm}
{1\over 2\alpha}
(k_\perp|Ue^{i{p_\perp^2\over\alpha s}y_\ast}(p_\perp^2\delta_j^k+2p_jp^k)U^\dagger|y_\perp)^{an}\simeq~{1\over 2\alpha}e^{i{k_\perp^2\over\alpha s}y_\ast-i(k,y)_\perp}
\Big\{
(k_\perp^2\delta_j^k+2k_jk^k)
\label{13.8}\\
&&\hspace{-12mm}
-{2gy_\ast\over\alpha s}(k_\perp^2\delta_j^k+2k_jk^k)k^l \partial_lU_yU^\dagger_y
+~2ig(\delta_j^k k^l-g^{kl}k_j-\delta_j^lk^k)
\partial_lU_yU^\dagger_y
\Big\}^{an}
\nonumber
\end{eqnarray}
It is easy to see now that the combination of formulas (\ref{13.6}) and (\ref{13.8}) (multiplied by $\theta(-y_\ast)$) proves Eq. (\ref{niid2pruv}) in the light-cone case.

\subsection{Shock-wave case}
If the characteristic transverse momenta of background field $l_\perp$ are of the same order of magnitude as the momenta 
of the ``quantum'' fields $p_\perp$ we have a ``shock-wave case'' when longitudinal size of background fields
$\sigma_\ast\sim~{\sigma s\over l_\perp^2}$ is much smaller than typical distances in quantum Feynman diagrams
$\sim~{\alpha s\over l_\perp^2}$ (recall that $\alpha\gg\sigma$). As in Ref. \cite{gTMD1}, we must consider separately two cases: $y_\ast$ inside and outside of the shock wave. The first case is simple:
since ${p_\perp^2\over\alpha s}y_\ast\sim {p_\perp^2\over\alpha s}\sigma_\ast\ll 1$ we can neglect 
$e^{{p_\perp^2\over\alpha s}y_\ast}$ factors in Eqs. (\ref{newlvertex}) and Eq. (\ref{lvertexcoord}) which effectively puts all operators on the light ray $y_\perp+{2\over s}z_\ast p_1$ so we return to the ``light-cone'' case considered  in the previous Section.

If $y_\ast$ is outside the shock wave, first we note that $\calo$ of Eq. (\ref{pa2}) can be replaced by pure gauge link
$[x_\ast,y_\ast]$. Indeed, let us compare the first and  the second terms in r.h.s. of Eq.  (\ref{pa2})
\begin{eqnarray}
&&\hspace{-11mm}
\calo_{\alpha}(x_\ast,y_\ast)~
=~[x_\ast,y_\ast]-{2ig\over\alpha s^2}\!\int_{y_\ast}^{x_\ast}\!\!\!dz_\ast
~\Big(z_\ast\{p^j,[x_\ast,z_\ast]F_{\bu j}(z_\ast)[z_\ast,y_\ast]\}
+...
\nonumber
\end{eqnarray}
The first term is $\sim 1$ while the second is 
$\sim {1\over\alpha s}\sigma_\ast p^j\partial_jU\sim {\sigma_\ast l_\perp^2\over\alpha s}\sim {\sigma\over\alpha}\ll 1$.
In a similar manner one can demonstrate that other terms in the r.h.s. of Eq.  (\ref{pa2}) are $\sim  {\sigma\over\alpha}$ 
in comparison to the first $[x_\ast,y_\ast]$ and therefore
the Lipatov vertex (\ref{newlvertex}) reduces to
\begin{eqnarray}
&&\hspace{-1mm}
L_j^{k,ab}(y_\perp,k_\perp; y_\ast)=~{1\over 2\alpha}
\Big[(k_\perp|[\infty,y_\ast]e^{i{p_\perp^2\over\alpha s}y_\ast}(p_\perp^2\delta_j^k+2p_jp^k)|y_\perp)
[y_\ast,\infty]_y
\nonumber\\
&&\hspace{-1mm}
-~{4\over s}(k_\perp|[\infty,y'_\ast]\!\int_{y_\ast}^\infty \! dy'_\ast~e^{i{p_\perp^2\over\alpha s}y'_\ast}p^k|y_\perp)
[y'_\ast,y_\ast]_yF_{\bu j}(y_\ast,y_\perp)[y_\ast,\infty]_y\Big]^{ab}
\nonumber\\
&&\hspace{-1mm}
=~{\theta(y_\ast)\delta^{ab}\over 2\alpha}(\delta^k_j p^2_\perp+2p_jp^k)e^{i{k_\perp^2\over\alpha s}y_\ast-i(k,y)_\perp}
+{\theta(-y_\ast)\over 2\alpha}
(k_\perp|Ue^{i{p_\perp^2\over\alpha s}y_\ast}(p_\perp^2\delta_j^k+2p_jp^k)U^\dagger|y_\perp)^{ab}
\nonumber\\
&&\hspace{-1mm}
-~2i{k^k\over k_\perp^2}e^{i{k_\perp^2\over\alpha s}y_\ast-i(k,y)_\perp}
[\infty,y_\ast]_yF_{\bu j}(y_\ast,y_\perp)[y_\ast,\infty]_y^{ab}
\label{newlvertex1}
\end{eqnarray}
because $[\infty,y_\ast]~=~\theta(-y_\ast)U+\theta(y_\ast)$ if $y_\ast$ is outside the shock wave.
Now we prove that the rest of r.h.s. of Eq. (\ref{lvertexcoord}) can be neglected 
\begin{eqnarray}
&&\hspace{-1mm}
-~{4ig\over\alpha s^2}(k_\perp^2\delta_j^k+2k_jk^k)k^l\!\int\!dz_\ast
~\big((z-y)_\ast\theta(z_\ast-y_\ast)+y_\ast\theta(-y_\ast)\big)[\infty,z_\ast]F_{\bu l}(z_\ast)[z_\ast,\infty]
\nonumber\\
&&\hspace{-1mm}
+~(\delta_j^k k^l-g^{kl}k_j-\delta_j^lk^k)
{4\over s}\!\int\!dz_\ast~[\theta(z_\ast-y_\ast)-\theta(-y_\ast)]
~[\infty,z_\ast]F_{\bu l}(z_\ast)[z_\ast,\infty]
~=~O\big(k_\perp^2{\sigma\over\alpha}\big)
\nonumber\\
\label{13.10}
\end{eqnarray}
To prove Eq. (\ref{13.10}) we first notice that at $y_\ast>0$ (and outside of the shock wave) the Eq. (\ref{13.10}) 
vanishes since $F_{\bu i}(z_\ast)=0$. 
Second, if $y_\ast<0$ the integral $\!\int_{y_\ast}^\infty\!dz_\ast~[\infty,z_\ast]F_{\bu l}(z_\ast)[z_\ast,\infty]$ can be replaced
by $\!\int_{-\infty}^\infty\!dz_\ast~[\infty,z_\ast]F_{\bu l}(z_\ast)[z_\ast,\infty]$ so Eq. (\ref{13.10}) reduces to
\begin{eqnarray}
&&\hspace{-1mm}
-~{4ig\over\alpha s^2}(k_\perp^2\delta_j^k+2k_jk^k)k^l\!\int\!dz_\ast
~z_\ast[\infty,z_\ast]F_{\bu l}(z_\ast)[z_\ast,\infty]
\label{13.11}
\end{eqnarray}
which is $\sim {k_\perp^2\over\alpha s}\sigma_\ast k^j\partial_jU\sim {k_\perp^4\over\alpha s}\sigma_\ast
\sim O\big(k_\perp^2{\sigma\over\alpha}\big)$. Now we see that the r.h.s of Eq. (\ref{newlvertex1}) coincides with the r.h.s. of Eq. (\ref{lvertexcoord}), so we have proved that Eq. (\ref{newlvertex}) agrees with Eq. (\ref{lvertexcoord}) with our accuracy 
$O\big({\sigma\over\alpha}\big)$. The last thing to note is that the integral of Eq. (\ref{lvertexcoord}) 
over $y_\ast$ with the weight ${2i\over s}e^{i\beta_B y_\ast}$ reproduces the Lipatov vertex (4.26) from
Ref. \cite{gTMD1}.

Finally, let us present the explicit form of the real (production) part of the kernel from Ref. \cite{gTMD1}
($\eta\equiv\ln\sigma$):
\begin{eqnarray}
&&\hspace{-1mm}
{d\over d\ln\sigma}\tilcaf_{(+\infty)i}^a(\beta_B, x_\perp) \calf_{(+\infty)j}^a(\beta_B, y_\perp)~
\label{realpart}\\
&&\hspace{-1mm}
\stackrel{\rm real}{=}~-\alpha_s{\rm Tr}\Big\{\!\int\!\dhd^2k_\perp
(x_\perp|\Big\{U{1\over\sigma\beta_Bs+p_\perp^2}
(U^\dagger k_k+p_kU^\dagger){\sigma \beta_Bsg_{\mu i}-2k^\perp_{\mu}k_i\over\sigma \beta_Bs+k_\perp^2}
\nonumber\\
&&\hspace{-1mm}
-~2k^\perp_\mu g_{ik}U{1\over \sigma\beta_Bs+p_\perp^2}U^\dagger -2g_{\mu k} \tilU{p_i\over\sigma\beta_Bs+p_\perp^2}
U^\dagger+{2k^\perp_\mu\over k_\perp^2}g_{ik}\Big\}
\ticalf^k_{(+\infty)}\big(\beta_B+{k_\perp^2\over\sigma s}\big)|k_\perp)
\nonumber\\
&&\hspace{5mm}
\times~(k_\perp|\calf^l_{(+\infty)}\big(\beta_B+{k_\perp^2\over\sigma s}\big)
\Big\{{\sigma \beta_Bs\delta^\mu_j-2k_\perp^{\mu}k_j\over\sigma\beta_Bs+k_\perp^2}
(k_lU+Up_l){1\over\sigma \beta_Bs+p_\perp^2}U^\dagger
\nonumber\\
&&\hspace{22mm}
-2k_\perp^\mu g_{jl}U{1\over \sigma\beta_Bs+p_\perp^2}U^\dagger
-~2\delta_l^\mu U{p_j\over\sigma\beta_Bs+p_\perp^2}U^\dagger
+2g_{jl}{k_\perp^\mu\over k_\perp^2}\Big\}|y_\perp)
~+~O(\alpha_s^2)
\nonumber
\end{eqnarray}
where
\begin{eqnarray}
&&\hspace{-0mm} 
\calf^{a\eta}_{(+\infty)i}(\beta_B,z_\perp)~\equiv~{2\over s}
\!\int\! dz_\ast ~e^{i\beta_B z_\ast} 
\big([\infty,z_\ast]_z^{am}gF^m_{\bu i}(z_\ast,z_\perp))^\eta,
\nonumber\\
&&\hspace{-0mm} 
\tilcaf^{a\eta}_{(+\infty)i}(\beta_B,z_\perp)~\equiv~{2\over s}
\!\int\! dz_\ast ~e^{-i\beta_B z_\ast} 
g\big(\tilF^m_{\bu i}(z_\ast,z_\perp)[z_\ast,\infty]_z^{ma}\big)^\eta
\end{eqnarray}
%


\end{document}